\documentclass[11pt]{article}
\pdfoutput=1
\usepackage{amsmath,amssymb,amsthm,bm,yhmath}
\usepackage{verbatim}
\usepackage{graphicx}
\usepackage{color}
\usepackage[round]{natbib}
\newcommand{\R}{\mathbb{R}}
\newcommand{\C}{\mathbb{C}}

\newcommand{\tr}{{\rm tr}}
\newcommand{\Sym}{{\rm Sym}}
\newcommand{\Diag}{{\rm Diag}}

\newcommand{\diag}{{\rm diag}}
\newcommand{\offdiag}{{\rm offdiag}}
\newcommand{\vecd}{{\rm vecd}}
\renewcommand{\vec}{{\rm vec}}
\newcommand{\Langle}{\langle\!\langle}
\newcommand{\Rangle}{\rangle\!\rangle}

\newcommand{\E}{{\rm E}}

\newcommand{\Cov}{{\rm Cov}}
\newcommand{\Exp}{{\rm Exp}}
\newcommand{\Log}{{\rm Log}}
\newcommand{\CR}{{\rm CR}}

\newtheorem{thm}{Theorem}%[section]
\newtheorem{prop}{Proposition}%[section]
\theoremstyle{definition}
\newtheorem{defn}{Definition}%[section]
\newtheorem{alg}{Algorithm}%[section]

\theoremstyle{remark}

\newcommand{\figurepath}{.}

\begin{document}

\title{Lognormal Distributions and Geometric Averages \\
of Positive Definite Matrices}
\author{Armin Schwartzman \\
Department of Statistics, North Carolina State University}
\date{\today}
\maketitle

\begin{abstract}
This article gives a formal definition of a lognormal family of probability distributions on the set of symmetric positive definite (PD) matrices, seen as a matrix-variate extension of the univariate lognormal family of distributions. Two forms of this distribution are obtained as the large sample limiting distribution via the central limit theorem of two types of geometric averages of i.i.d. PD matrices: the log-Euclidean average and the canonical geometric average. These averages correspond to two different geometries imposed on the set of PD matrices. The limiting distributions of these averages are used to provide large-sample confidence regions for the corresponding population means. The methods are illustrated on a voxelwise analysis of diffusion tensor imaging data, permitting a comparison between the various average types from the point of view of their sampling variability.
\end{abstract}

%========================================================================
\section{Introduction}
\label{sec:Introduction}

A scalar random variable is called lognormal if its logarithm is normally distributed. Analogously, a random positive definite (PD) matrix may be called lognormal if its matrix logarithm, a random symmetric matrix, has a matrix-variate normal distribution. The resulting distribution on the set of PD matrices can be naturally called the PD-matrix-variate lognormal distribution.

While useful as a probability model for PD matrix data \citep{Schwartzman:Thesis2006}, it is argued here that this distribution plays a more central role in multivariate statistics. Just like the scalar lognormal distribution may be obtained as the limiting distribution of the geometric average of i.i.d. random variables \citep{Johnson:1970}, it is shown here that two types of PD-matrix-variate lognormal distributions arise as the limiting distributions of two types of geometric averages of i.i.d. random PD matrices. The main goal of this paper is to justify and make explicit the definition of such distributions. Moreover, it is shown how these distributions can be used to construct confidence regions (CR's) for those geometric averages in data analysis.

The first geometric average considered in this paper, called log-Euclidean average, is obtained by taking the average of the PD matrices after a matrix-log transformation, and transforming the result back to the original space \citep{Arsigny:2006,Schwartzman:Thesis2006}. Application of the multivariate central limit theorem (CLT) in the log domain leads directly to the definition of a PD-matrix-variate lognormal distribution, which we call Type I, so that the log-Euclidean average can be regarded as approximately lognormal for large sample sizes. The second geometric average, called here canonical geometric average, is the Fr\'{e}chet average associated with the so-called canonical, affine-invariant or generalized Frobenius Riemannian metric defined on the set of PD matrices \citep{Kobayashi:1996,Lang:1999,Moakher:2005,Smith:2005,Schwartzman:Thesis2006,Lenglet:2006,Fletcher:2007,Zhu:2009}. This case requires application of a Riemannian manifold version of the CLT, and the limiting distribution leads to the definition of a second type of lognormal distribution, called here Type II. The Type I and Type II distributions coincide when the mean parameter in both is a multiple of the identity matrix.

As a side benefit, the theoretical analysis above allows establishing some properties of these geometric averages and their corresponding geometric population means, such as uniqueness, consistency and invariance. The two geometric averages are constrasted with the simpler arithmetic or Euclidean average of PD matrices, which is also consistent but asymptotically normal. Their different properties are described in terms of the geometry of the set of PD matrices.

In classical multivariate statistics, the most common probability model for PD matrices has been the Wishart distribution. Related parametric models for PD matrices include the multivariate gamma, inverse Wishart, and inverse multivariate gamma, as well as noncentral versions of these distributions \citep{Anderson:2003,Gupta:2000,Muirhead:1982}. These models assume that the observations are real-valued multivariate vectors and have as a goal to model their population or sample covariance matrix. The idea of using a matrix logarithm of PD matrices to flexibly model covariance matrices in a similar multivariate context was explored by \citet{Leonard:1992} and \citet{Chiu:1996}.

Today, new data exist where the observations themselves (after pre-processing) take the form of PD matrices. In Diffusion Tensor Imaging (DTI), a special modality of magnetic resonance imaging, each voxel contains a random $3 \times 3$ PD matrix called diffusion tensor \citep{Basser:1996,LeBihan:2001,Bammer:2002}. In the brain, the diffusion tensor describes the local spatial pattern of water diffusion, its eigenvalues related to the type of tissue and its health, and its eigenvectors related to the underlying spatial direction of neural fibers. In tensor-based morphometry (TBM), brain images are spatially registered to a template and the transformation is summarized at each voxel by a $3 \times 3$ PD matrix called deformation tensor \citep{Joshi:2007,Lepore:2008}.

Both the log-Euclidean and canonical geometric averages of PD matrices, as well as the simpler Euclidean average, have been used for DTI data \citep{Arsigny:2006,Schwartzman:Thesis2006,Fletcher:2007,Zhu:2009,Schwartzman:2010a} and TBM data \citep{Lepore:2008}. The results in this paper enable the construction of CR's for these averages. Interesting questions arise about how to visualize these 6-dimensional regions, where the elements of these regions and the averages themselves are 6-dimensional objects represented as 3-dimensional ellipsoids. In order to capture the largest variation inside the CR, here we consider the display of the two ellipsoids at the boundary of the CR along its first principal axis. Interestingly, finding such boundary points for the canonical geometric average requires solving a special case of the continuous-time algebraic Ricatti equation, whose solution is in itself a form of canonical geometric average.

As a specific data example, the three types of average are compared in an analysis of DTI images corresponding to a group of 34 10-year-old children \citep{Dougherty:2007,Schwartzman:2010a}. The analysis shows that the log-Euclidean and canonical geometric averages are statistically indistinguishable, whereas the Euclidean and log-Euclidean are within their respective ranges of sampling variability for this sample size for most voxels. These results suggest that the theoretical debate about the choice of average \citep{Whitcher:2007,Dryden:2009,Pasternak:2010} may sometimes be of no practical importance, making the Euclidean or log-Euclidean preferable for their computational simplicity.

The parametric approach followed in this paper contrasts and complements the nonparametric approach of \citet{Osborne:2013,Ellingson:2013}, which uses bootstrapping to establish significance in two-sample tests of canonical geometric means but is restricted to analysis of a single voxel because of computational complexity. In contrast, the parametric approach followed here can be easily applied in parallel to thousands of voxels in the entire brain volume.

The rest of the paper is organized as follows. Section \ref{sec:lognormal} introduces the Euclidean and log-Euclidean averages, as well as the lognormal distribution of Type I. Section \ref{sec:canonical} presents the canonical geometric average and the resulting lognormal distribution of Type II. Section \ref{sec:dataExample} presents the DTI example, concluding with a discussion in Section \ref{sec:discussion}. Relevant auxiliary material appear in Appendix \ref{sec:basics} and proofs are placed in Appendix \ref{sec:proofs}.

%===================================================================
\section{The log-Euclidean average and the PD-matrix lognormal distribution of Type I}
\label{sec:lognormal}

%-------------------------------------------------------------------
\subsection{The set of PD matrices}

\begin{figure}[t!]
  \centering
  \includegraphics[width=\textwidth]{\figurepath/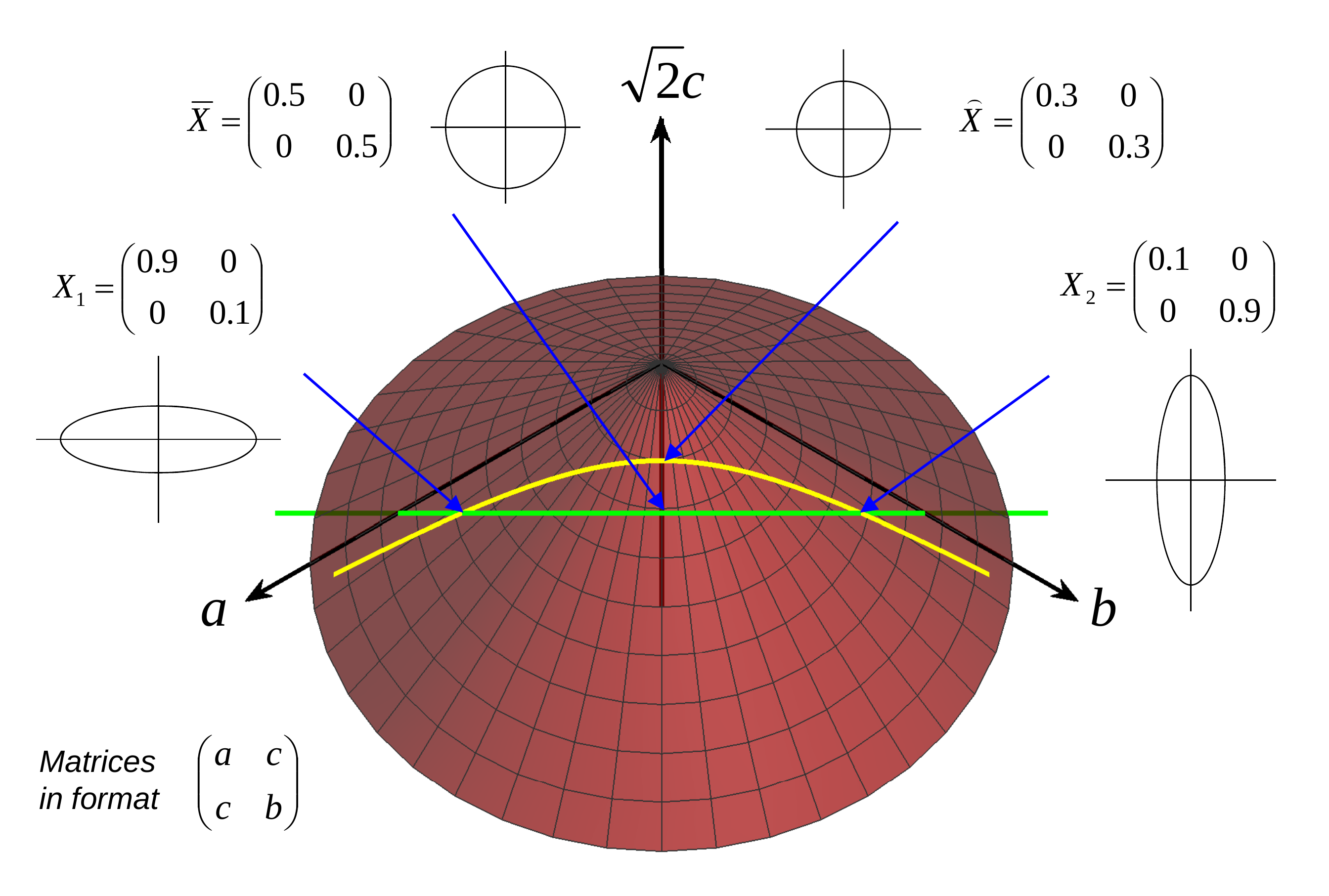}
  \caption{A $2 \times 2$ matrix $X$ with diagonals $a$, $b$ and off-diagonal $c$ is PD if and only if $a>0$, $b>0$ and $ab-c^2>0$. The set of valid triplets $\vecd(X) = (a,b,\sqrt{2} c)'$ (see also Appendix \ref{sec:vecd}) is an open cone in $\R^3$. For example, the diagonal matrices $X_1 = \Diag(0.9,0.1)$ and $X_2 = \Diag(0.1,0.9)$ are represented by the triplets $\vecd(X_1)=(0.9,0.1,0)'$ and $\vecd(X_2)=(0.1,0.9,0)'$ respectively.
Because this set is convex, interpolation between any two points is permitted, and the Euclidean average $\bar{X} = \Diag(0.5,0.5)$ is PD.
Extrapolation, however, might result in matrices that are not PD: the green straight line connecting $X_1$ and $X_2$ extends beyond the boundaries of the cone. The yellow hyperbola connecting $X_1$ and $X_2$ is the matrix exponential of the straight line between $\log X_1$ and $\log X_2$, always inside the cone by construction. The midpoint along this line is the log-Euclidean average \eqref{eq:log-Euclidean-avg} equal to $\wideparen{X} = \Diag(0.3,0.3)$. In this case, it also coincides with the canonical geometric average defined via \eqref{eq:canonical-geometric-avg}.
\label{fig:cone}}
\end{figure}

It is useful to think of the set $\Sym^+(p)$ of $p \times p$ real PD matrices in relation to the set $\Sym(p)$ of $p \times p$ real symmetric matrices in three different ways. First, $\Sym^+(p)$ is a simply connected open subset of $\Sym(p)$, whose boundary is the set of $p \times p$ symmetric matrices with one or more zero eigenvalues. It is also a convex cone in $\Sym(p)$ because $X_1, X_2 \in \Sym^+(p)$ implies $a_1 X_1 + a_2 X_2 \in \Sym^+(p)$ for any scalars $a_1, a_2 > 0$. However, it is not a vector space; e.g. the negative of a PD matrix and some linear combinations of PD matrices are not PD. Figure \ref{fig:cone} shows the shape of the cone for $p=2$ by embedding $\Sym^+(2)$ in $\R^3$. % (see also Section \ref{sec:vecd}).

Second, $\Sym^+(p)$ and $\Sym(p)$ have a one-to-one correspondence via the matrix exponential and logarithm
\begin{equation}
\label{eq:diffeomorphism}
\Sym^+(p) = \exp\Big[\Sym(p)\Big], \qquad
\Sym(p) = \log\Big[\Sym^+(p)\Big],
\end{equation}
(see Appendix \ref{sec:exp-log}). Third, $\Sym^+(p)$ is a differentiable manifold whose tangent space at any point can be identified as a copy of $\Sym(p)$. Furthermore, it can be given a Riemannian manifold structure by defining a metric, which in this paper will be Euclidean, log-Euclidean, or canonical (affine-invariant) (see Appendix \ref{sec:Euclidean-geom} and \ref{sec:canonical-geom}). These relationships will play various roles in the development below.

A useful geometrical representation of the elements of $\Sym^+(p)$ themselves is the association of $X \in \Sym^+(p)$ with the ellipsoid in $\R^p$
\begin{equation}
\label{eq:ellipsoid}
\mathcal{E}(X) = \{w \in \R^p: w'X^{-1} w = 1\}.
\end{equation}
This ellipsoid can be thought of as an isosurface of a multivariate normal distribution in $\R^p$ with covariance $X$. This representation is useful for visualization of $X$ as the semiaxes of $\mathcal{E}(X)$ correspond to the eigenvectors of $X$ scaled by the square root of the eigenvalues. The representation of some elements of $\Sym(2)$ as ellipses is illustrated in Figures \ref{fig:cone} and \ref{fig:random-ellipses}.

%-------------------------------------------------------------------
\subsection{The Euclidean average of PD matrices and the symmetric-matrix normal distribution}
\label{sec:Euclidean}

Before presenting the log-Euclidean average, it is helpful to first consider the Euclidean average for comparison. Given PD matrices $X_1,\dots,X_n \in \Sym^+(p)$, the convexity of $\Sym^+(p)$ implies that the Euclidean average
\begin{equation}
\label{eq:Euclidean-avg}
\bar{X} = \frac{1}{n} \sum_{i=1}^n X_i
\end{equation}
is also an element of $\Sym^+(p)$ (Figure \ref{fig:cone}). Suppose $X_1,\dots,X_n \in \Sym^+(p)$ are pairwise i.i.d. such that $\E(X_i) = M \in \Sym^+(p)$ (capital $\mu$) and $\Cov(\vecd(X_i)) = \Sigma \in \Sym^+(q)$. Here, $\vecd(\cdot)$ can be in general any vectorization operator, but we prefer the specific operator defined by \eqref{eq:vecd} in Appendix \ref{sec:vecd}. Then by the multivariate CLT, as $n \to \infty$,
\begin{equation}
\label{eq:Euclidean-avg-CLT}
\bar{X} \rightarrow M, \qquad
\sqrt{n}\vecd\big(\bar{X} - M\big) \Rightarrow N(0,\Sigma)
\end{equation}
in probability and in distribution, respectively. Thus $\bar{X} \in \Sym^+(p)$ is a consistent estimator of $M$ and, for large $n$, is approximately normally distributed in $\Sym(p)$ with mean $M$ and covariance $\Sigma/n$ ($\Sigma$ can be estimated consistently by the sample covariance matrix of $\vecd(X_1),\dots,\vecd(X_n)$ if finite fourth order moments are also assumed).

In this first view, the PD matrices $X_1,\dots,X_n \in \Sym^+(p)$ are treated as if they were mere symmetric matrices. As $n$ increases, the distribution of $\bar{X}$ becomes more concentrated around $M$ and the PD constraints at the boundary of $\Sym^+(p)$ become asymptotically irrelevant. However, since the normal distribution is not defined on $\Sym^+(p)$ for any $n$ because of the boundary constraints, the asymptotic normal distribution of $\bar{X}$ is technically defined on the tangent space to $\Sym^+(p)$ at $M$, equal to $\Sym(p)$. With some abuse of notation, we write this as
\begin{equation}
\label{eq:Euclidean-avg-approx}
\bar{X} \mathop{\sim}^{.} N(M,\Sigma/n),
\end{equation}
where we use the notational shorthand
\begin{equation}
\label{eq:matrix-normal}
X \sim N(M,\Sigma) \qquad \Leftrightarrow \qquad \vecd(X) \sim N(\vecd(M),\Sigma).
\end{equation}
The specific choice of the vectorization operator $\vecd(\cdot)$ given by \eqref{eq:vecd} gives an elegant definition of the symmetric-matrix-variate normal distribution above \citep{Schwartzman:2008c}, which in the case of identity covariance reduces to the random matrix model called Gaussian Orthogonal Ensemble \citep{Mehta:1991}.

Based on \eqref{eq:Euclidean-avg-CLT}, a confidence region (CR) for $M$ of asymptotic level $\alpha$ is given by the $q$-dimensional ellipsoid
\begin{equation}
\label{eq:Euclidean-avg-CR}
\CR_{n,\alpha}(\bar{X}) = \left\{M \in \Sym^+(p): ~ n \vecd(\bar{X} - M)' \hat{\Sigma}^{-1} \vecd(\bar{X} - M) \le \chi^2_{q,1-\alpha}\right\},
\end{equation}
where $\chi^2_{q,1-\alpha}$ is the $1-\alpha$ quantile of the $\chi^2$ distribution with $q$ degrees of freedom and $\hat{\Sigma}$ is a consistent estimator of $\Sigma$ such as the sample covariance matrix of $\vecd(X_1),\dots,\vecd(X_n)$.

%-------------------------------------------------------------------
\subsection{The log-Euclidean average of PD matrices}
\label{sec:log-Euclidean-avg}

A second view of $\Sym^+(p)$ is as the image of $\Sym(p)$ via the matrix exponential. Just like positive scalar data is sometimes log-transformed before analysis, the analogous approach to analysis of PD matrices is to transform them into symmetric matrices by a matrix log transformation \citep{Arsigny:2006,Schwartzman:Thesis2006,Fletcher:2007,Zhu:2009,Schwartzman:2010a,Lepore:2008}. Given PD matrices $X_1,\dots,X_n \in \Sym^+(p)$, mapping them to  $\Sym(p)$ via the matrix logarithm, computing their Euclidean average there and then mapping the result back to $\Sym^+(p)$ results in the log-Euclidean average
\begin{equation}
\label{eq:log-Euclidean-avg}
\wideparen{X} = \exp \bigg( \frac{1}{n} \sum_{i=1}^n \log X_i \bigg).
\end{equation}
The symbol $\wideparen{X}$ reflects that this is a ``curved'' version of the Euclidean average (Figure \ref{fig:cone}).

Given a random PD matrix $X$ with probability measure $Q$ on $\Sym^+(p)$, we can define the log-Euclidean mean
\begin{equation}
\label{eq:log-Euclidean-mean}
M = \exp \left[ \E(\log X) \right],
\end{equation}
where the expectation is taken with respect to the measure $Q$, so that the log-Euclidean average \eqref{eq:log-Euclidean-avg} corresponds to the empirical measure $\hat{Q}_n = (1/n) \sum_{i=1}^n \delta(X_i)$.
A direct consequence of the analytical property \eqref{eq:analytical} in Appendix \ref{sec:exp-log} is that the log-Euclidean mean and the log-Euclidean average are equivariant under similarity transformations and matrix inversion. Let $GL(p)$ denote the set of $p \times p$ real invertible matrices (general linear group). If $X$ is a random PD matrix with log-Euclidean mean $M$, then:
\begin{enumerate}
\item The log-Euclidean mean of $A X A^{-1}$ is $A M A^{-1}$, for $A \in GL(p)$.
\item The log-Euclidean mean of $X^{-1}$ is $M^{-1}$.
\end{enumerate}
Similarly, given PD matrices $X_1,\ldots,X_n$ with log-Euclidean average \eqref{eq:log-Euclidean-avg}:
\begin{enumerate}
\item The log-Euclidean average of $A X_1 A^{-1}, \ldots, A X_n A^{-1}$ is $A \wideparen{X} A^{-1}$, for $A \in GL(p)$.
\item The log-Euclidean average of $X_1^{-1}, \ldots, X_n^{-1}$ is $\wideparen{X}^{-1}$.
\end{enumerate}

%-------------------------------------------------------------------
\subsection{The log-Euclidean average and the PD-matrix lognormal distribution of Type I}
\label{sec:log-Euclidean-distr-I}
Suppose now that $X_1,\dots,X_n \in \Sym^+(p)$ are pairwise i.i.d. PD matrices such that $\E(\log X_i) = \log M \in \Sym(p)$ and $\Cov(\vecd(\log X_i)) = \Sigma \in \Sym^+(q)$. Let $\wideparen{X}$ denote their log-Euclidean average, given by \eqref{eq:log-Euclidean-avg}. Then, applying the multivariate CLT to the transformed matrices $\log X_1,\dots,\log X_n$ gives that, as $n \to \infty$,
\begin{equation}
\label{eq:log-Euclidean-avg-CLT}
\wideparen{X} \rightarrow M, \qquad
\sqrt{n}\vecd\big(\log\wideparen{X} - \log M\big) \Rightarrow N(0,\Sigma)
\end{equation}
in probability and in distribution, respectively. Thus $\wideparen{X} \in \Sym^+(p)$ is a consistent estimator of the log-Euclidean population mean \eqref{eq:log-Euclidean-mean} ($\Sigma$ can be estimated consistently by the sample covariance matrix of the vectors $\vecd(\log X_1),\ldots,\vecd(\log X_n)$ if finite fourth order moments are also assumed).

More interestingly, the log-Euclidean average $\wideparen{X}$ is asymptotically lognormal in the sense that, for large $n$, $\log\wideparen{X}$ is approximately normally distributed in $\Sym(p)$ with mean $\log M$ and covariance $\Sigma/n$. This leads to the first definition of the lognormal distribution.

\begin{defn}
\label{defn:log-normal-distr-I}
We say that $X \in \Sym^+(p)$ has a {\em PD-matrix-variate lognormal distribution of Type I} with parameters $M \in \Sym^+(p)$ and $\Sigma \in \Sym^+(q)$, denoted $X \sim LN_{I}(M,\Sigma)$, if $\log X \sim N(\log M,\Sigma)$.
\end{defn}

Definition \ref{defn:log-normal-distr-I} has the interpretation that the parameter $M$ is precisely the log-Euclidean mean \eqref{eq:log-Euclidean-mean} of the lognormal variable $X$. Based on \eqref{eq:log-Euclidean-avg-CLT}, we can write the approximate distribution of the log-Euclidean average for large $n$ as
\begin{equation}
\label{eq:log-Euclidean-avg-approx}
\wideparen{X} \mathop{\sim}^{.} LN_{I}(M,\Sigma/n).
\end{equation}

\begin{figure}
  \centering
  a\includegraphics[width=2.3in]{\figurepath/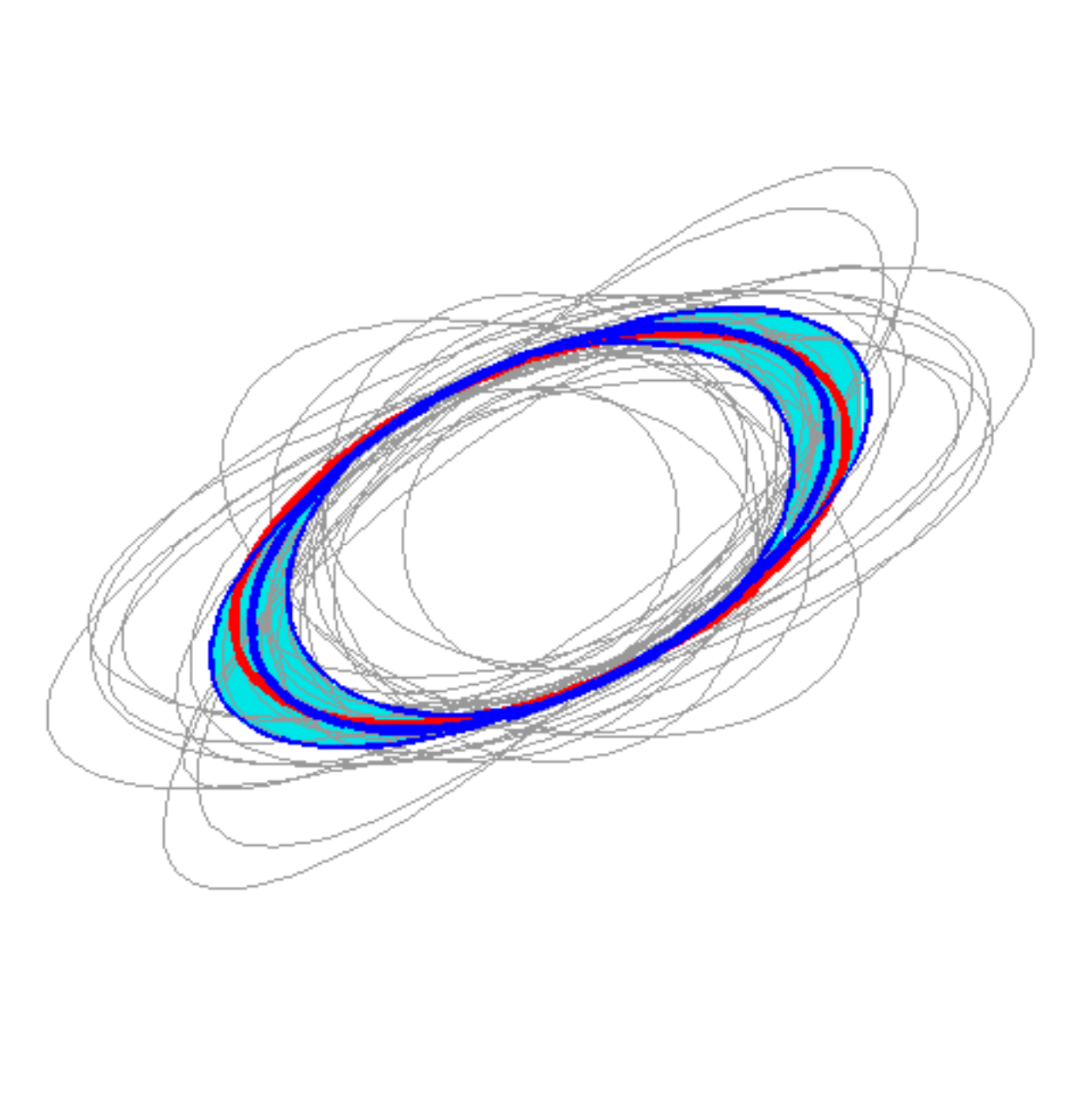}
  b\includegraphics[width=2.3in]{\figurepath/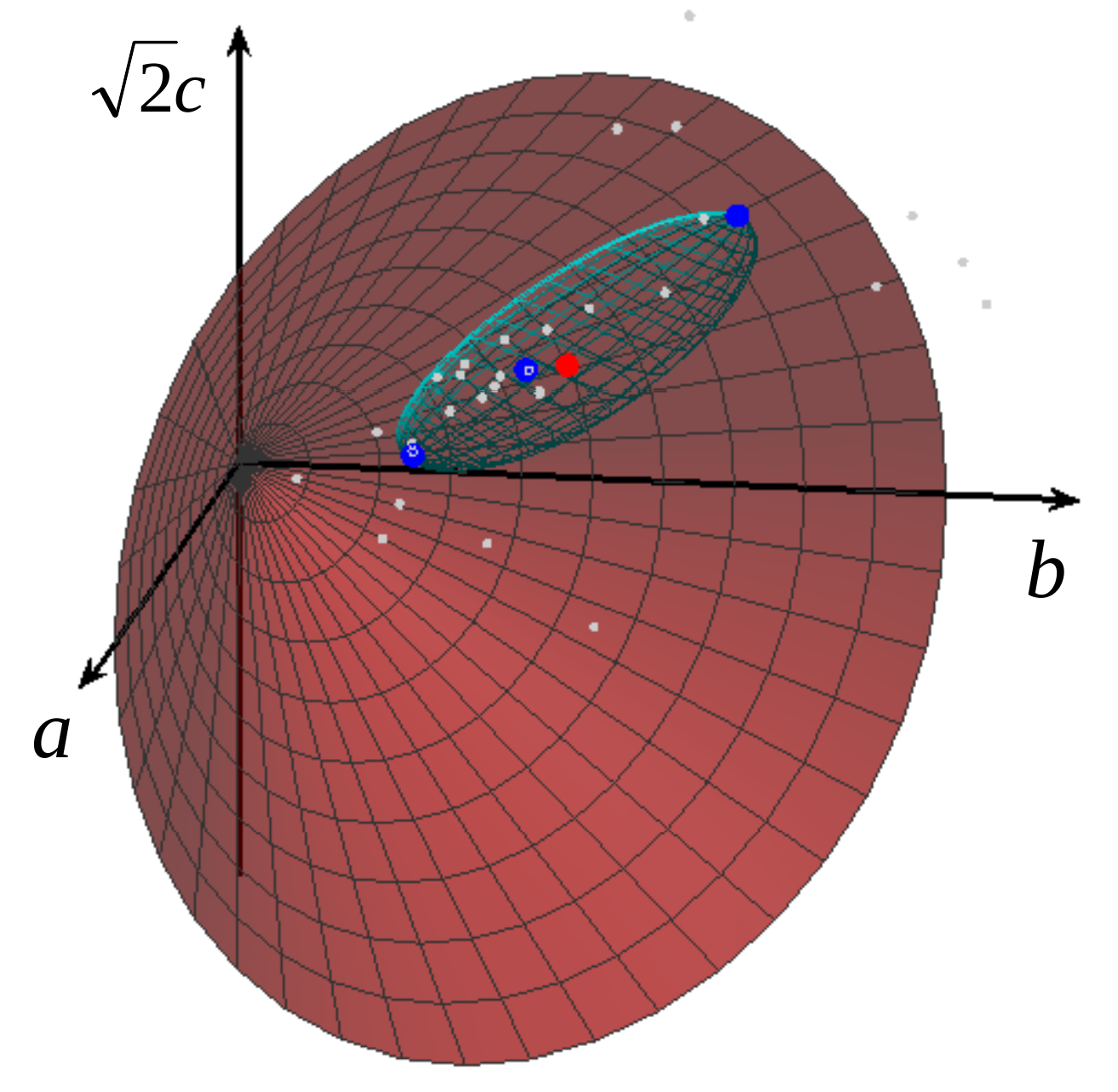}
  \caption{(a) 30 samples (gray) generated from a lognormal distribution $X \sim LN(M,\Sigma)$ with parameters $M = \Diag(4,1)$ and $\Sigma = \Diag(0.5,0.5,0)$ represented by ellipses via \eqref{eq:ellipsoid}. Notice the variation around the mean ellipse $M$ (red) with semiaxes $\sqrt{4}=2$ and $\sqrt{1}=1$. Superimposed are the log-Euclidean average (inner blue) and two exreme points of the 95\% CR (outer blue). (b) Same as in (a) displayed in the coordinates of Figure \ref{fig:cone}, where each ellipse in (a) corresponds to a point of the same color inside the cone. The 95\% CR is represented by the cyan mesh.}
  \label{fig:random-ellipses}
\end{figure}

To generate random lognormal PD matrices $X \sim LN_{I}(M,\Sigma)$, the recipe is to generate random normal symmetric matrices $Y \sim N(\log M,\Sigma)$ and then apply the matrix exponential $X = \exp(Y)$. An illustration for $p=2$ is shown in Figure \ref{fig:random-ellipses}. The representation of the random matrices and their log-Euclidean mean and average as ellipses via \eqref{eq:ellipsoid} nicely visualizes their positive definiteness and variability. However, it is less useful for representing CR's, as explained below.

The following result helps establish the density of the PD-matrix-variate lognormal distribution of Type I. The proof is given in Appendix \ref{sec:proofs}.

\begin{thm}
\label{thm:log-density}
Let $\lambda_1 \ge \ldots \ge \lambda_p > 0$ denote the eigenvalues of $X \in \Sym^+(p)$. The Jacobian of the transformation $Y = \log X \in \Sym(p)$ is given by
\begin{equation}
\label{eq:Jacobian}
J(X) = \mathcal{J}(Y \to X) = \frac{1}{\lambda_1 \ldots \lambda_p}
\prod_{i<j} g(\lambda_i,\lambda_j),
\end{equation}
where
\begin{equation}
g(\lambda_i,\lambda_j) = \begin{cases}
\left(\log \lambda_i - \log \lambda_j\right)/\left(\lambda_i - \lambda_j\right), & \lambda_i > \lambda_j \\
1/\lambda_i, & \lambda_i = \lambda_j. \end{cases}
\end{equation}
\end{thm}

Making the change of variable $Y = \log X$ in Definition \ref{defn:log-normal-distr-I} and applying Theorem \ref{thm:log-density} gives that the density of $X \sim LN_{I}(M,\Sigma)$ is
\begin{equation}
\label{eq:lognormal-density-I}
f(X; M, \Sigma) = \frac{J(X)}{(2\pi)^{q/2} |\Sigma|^{1/2}}
\exp \biggl(-\frac{1}{2} \vecd(\log X - \log M)' \Sigma^{-1}
\vecd(\log X - \log M)\biggr)
\end{equation}
with respect to Lebesgue measure on $\Sym^+(p)$, where $J(X)$ is the Jacobian \eqref{eq:Jacobian}. An interesting special case occurs when $M = a I_p$, $a > 0$, and $\Sigma$ is an orthogonally invariant covariance matrix with respect to the operator $\vecd(\cdot)$ \citep{Schwartzman:2008c}, e.g. $\Sigma = I_q$. In that case the density \eqref{eq:lognormal-density-I} becomes  a function of the eigenvalues of $X$ only and is orthogonally invariant in the sense that $f(X) = f(Q X Q')$ for any $p\times p$ orthogonal matrix $Q$. We omit the explicit expression of the density in that case for brevity.
\begin{comment}
\[
f(X; I_p, I_q) = \frac{1}{(2\pi)^{q/2}}
\frac{1}{\lambda_1 \ldots \lambda_p}
\prod_{i<j} \frac{\log \lambda_j - \log \lambda_i}{\lambda_j - \lambda_i}
\cdot
\exp \left[-\frac{1}{2} \sum_{i=1}^p (\log \lambda_i)^2\right]
\]
\end{comment}

\begin{figure}[t]
  \centering
  a\includegraphics[height=2in,trim=0.5in 0 0.5in 0,clip]{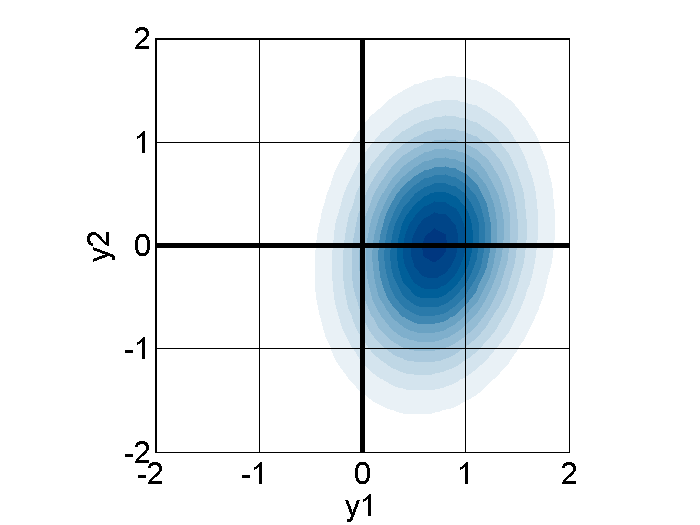}\includegraphics[height=2in,trim=0.5in 0 0.5in 0,clip]{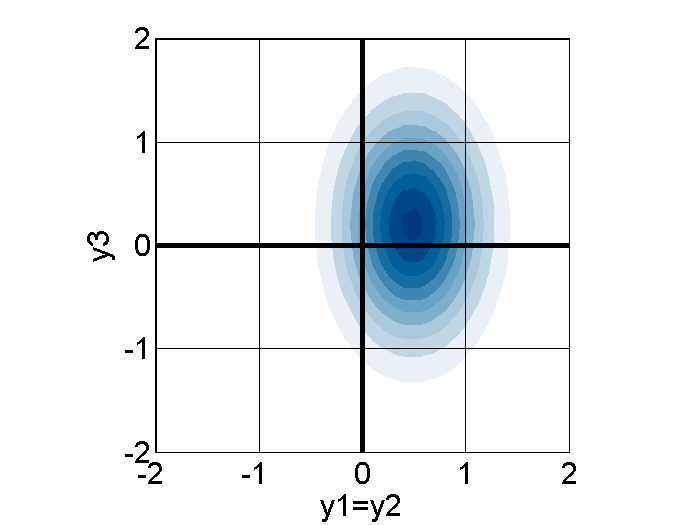}
  b\includegraphics[height=2in,trim=0.5in 0 0.5in 0,clip]{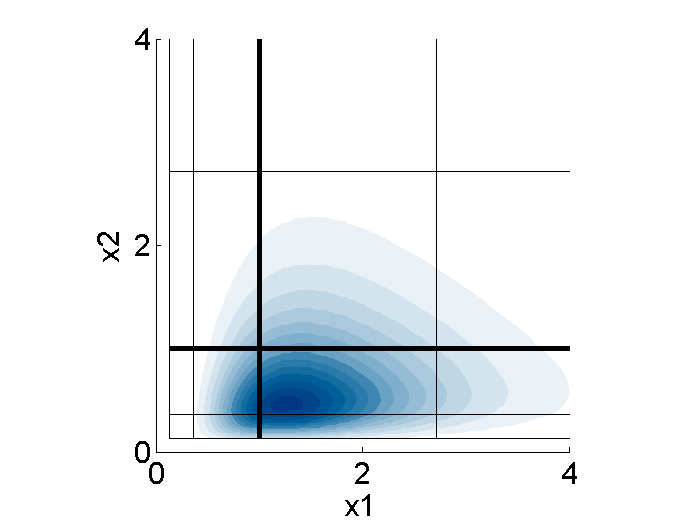}\includegraphics[height=2in,trim=0.5in 0 0.5in 0,clip]{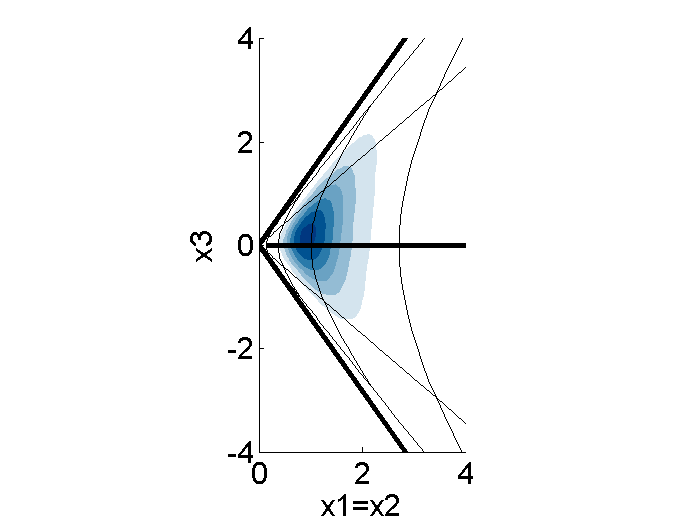}
  \caption{(a) Density of $y = \vecd(Y) \sim N(\vecd(\log M), \Sigma)$ with parameters \eqref{eq:params-I} displayed in two planes: $y_3 = 0$ (left column) and $y_1 = y_2$ (right column). The density contours increase in height for darker colors. (b) Density of $x = \vecd(X)$ with $X = \exp(Y)$ displayed in two planes: $x_3 = 0$ (left column) and $x_1 = x_2$ (right column).}
\label{fig:lognormal-I-density}
\end{figure}

As an example of what the lognormal density looks like, Figure \ref{fig:lognormal-I-density} shows the construction of a lognormal density of Type I in $\Sym^+(2)$ with parameters
\begin{equation}
\label{eq:params-I}
M = \exp\begin{pmatrix} 0.7 & 0.2 \\ 0.2 & 0 \end{pmatrix} = \begin{pmatrix} 2.05 & 0.29 \\ 0.29 & 1.03 \end{pmatrix}, \qquad
\Sigma = \begin{pmatrix} 0.25 & 0.05 & 0 \\ 0.05 & 0.5 & 0 \\ 0 & 0 & 0.5 \end{pmatrix}.
\end{equation}
Panel (a) shows two views of the density of the normal vector $y = \vecd(Y) \sim N(\vecd(\log M), \Sigma)$. Applying the transformation $X = \exp(Y)$ yields the density of the vector $x = \vecd(X)$, shown in two views in Panel (b). The gridlines in Panel (a) have been mapped directly to Panel (b) in order to illustrate the deformative effect of the matrix exponential. Notice the shape of the $\Sym^+(2)$ cone on the right column.

Finally, another nice property of the lognormal distribution of Type I is that, if $X_1,\dots,X_n$ are i.i.d. $LN(M,\Sigma)$, then the MLEs of $M$ and $\Sigma$ are given, respectively, by the log-Euclidean average and sample covariance matrix
$$
\hat{M} = \wideparen{X} = \exp \bigg( \frac{1}{n} \sum_{i=1}^n \log X_i \bigg),
\qquad
\hat{\Sigma} = \frac{1}{n} \sum_{i=1}^n \vecd(\log X_i) \vecd(\log X_i)'.
$$

%-------------------------------------------------------------------
\subsection{A confidence region for the log-Euclidean mean}
\label{sec:log-Euclidean-CR}

Based on \eqref{eq:log-Euclidean-avg-approx}, a CR for the log-Euclidean mean of asymptotic level $\alpha$ is given by
\begin{equation}
\label{eq:log-Euclidean-avg-CR}
\CR_{\rm n,\alpha}(\wideparen{X}) = \left\{M \in \Sym^+(p): ~ n \vecd(\log \wideparen{X} - \log M)' \hat{\Sigma}^{-1} \vecd(\log \wideparen{X} - \log M) \le \chi^2_{q,1-\alpha}\right\},
\end{equation}
where $\chi^2_{q,1-\alpha}$ is the $1-\alpha$ quantile of the $\chi^2$ distribution with $q$ degrees of freedom and $\hat{\Sigma}$ is a consistent estimator of $\Sigma$ such as the sample covariance matrix of $\vecd(\log X_1),\dots,\vecd(\log X_n)$.

An example of such a CR is shown in Figure \ref{fig:random-ellipses}b. Note that the 3-dimensional ellipsoid determined by \eqref{eq:log-Euclidean-avg-CR} in the log domain gets distorted after inverting the log into the original domain, so that the log-Euclidean average does not correspond to the Euclidean center of the CR but rather to its log-Euclidean center. The points on the CR mesh were obtained by solving for $M$ in \eqref{eq:log-Euclidean-avg-CR} using the principal components of $\hat{\Sigma}$ as basis. Particularly, marked in blue are the two extreme points along the first principal component of the CR (computed in the log domain), computed by inverting \eqref{eq:log-Euclidean-avg-CR} according to
\begin{equation*}
\label{eq:log-Euclidean-avg-CR-PCA}
\exp\left[\log\wideparen{X} \pm \vecd^{-1}\left( \lambda_1 V_1 \chi^2_{q,1-\alpha}/n \right)\right],
\end{equation*}
where $\lambda_1$, $V_1$ are the first eigenvalue and first eigenvector of the covariance matrix $\hat{\Sigma}$. These two points are also represented as ellipses in Figure \ref{fig:random-ellipses}a. It is worth emphasizing that the shaded region in Figure \ref{fig:random-ellipses}a is not a CR because it only captures the variability along the first principal component. While the true mean is not contained in the shaded region in Figure \ref{fig:random-ellipses}a, it is indeed contained in the CR in Figure \ref{fig:random-ellipses}b. Other visualization schemes may also be devised, for instance, by choosing points on the boundary of the CR that are multiples of the log-Euclidean average or whose corresponding ellipses have the largest and smallest plotted area.

%===================================================================
\section{The canonical geometric average and the PD-matrix lognormal distribution of Type II}
\label{sec:canonical}

A second type of geometric average, called here canonical geometric average, is obtained as the Fr\'{e}chet average according to the so-called canonical or affine-invariant metric defined on $\Sym^+(p)$ \citep{Arsigny:2006,Lenglet:2006,Fletcher:2007}. The canonical geometric average also leads asymptotically to a lognormal distribution, but of a different kind.

%-------------------------------------------------------------------
\subsection{The canonical geometric mean in $\Sym^+(p)$}
\label{sec:canonical-mean}

In general, the Fr\'{e}chet mean of a probability measure $Q$ on a complete metric space with distance $\rho$ is the minimizer of the function $F(P) = \int \rho^2(P,X) \,Q(dX)$ \citep{Frechet:1948,Pennec:1999,Bhattacharya:2003,Bhattacharya:2005}. In particular, for the empirical measure $\hat{Q}_n = (1/n) \sum_{i=1}^n \delta(X_i)$ of a set of i.i.d. data points $X_1,\ldots,X_n$, the minimizer of the function $\hat{F}_n(P) = (1/n) \sum_{i=1}^n \rho^2(P,X_i)$ is called the sample Fr\'{e}chet mean or Fr\'{e}chet average.

Seen in this context, the Euclidean average \eqref{eq:Euclidean-avg} and corresponding Euclidean mean $\E(X) = \int X \,Q(dX)$ are respectively the unique Fr\'{e}chet average and mean according to the Euclidean metric defined on $\Sym^+(p)$ and given by the Frobenius distance defined via %$d_{\rm E}^2(P,X) = |X-P|^2 = \tr[(X-P)^2]$.
\begin{equation}
\label{eq:Euclidean-distance}
d_{\rm E}^2(P,X) = |X-P|^2 = \tr[(X-P)^2].
\end{equation}
Similarly, the log-Euclidean average \eqref{eq:log-Euclidean-avg} and log-Euclidean mean \eqref{eq:log-Euclidean-mean} are respectively the unique Fr\'{e}chet average and mean according to a log-Euclidean distance defined on $\Sym^+(p)$ defined via \citep{Arsigny:2006} %$d_{\rm LE}^2(P,X) = |\log X - \log P|^2 = \tr[(\log X - \log P)^2]$ \citep{Arsigny:2006}.
\begin{equation}
\label{eq:log-Euclidean-distance}
d_{\rm LE}^2(P,X) = |\log X - \log P|^2 = \tr[(\log X - \log P)^2].
\end{equation}

If the measure $Q$ is defined on a manifold embedded in Euclidean space, then the Fr\'{e}chet mean associated with the restriction to the manifold of the Euclidean distance in the ambient space is called the extrinsic mean. Because of the inclusion $\Sym^+(p) \subset \Sym(p)$, the embedding of $\Sym^+(p)$ in Euclidean space is trival with no change in dimension. Moreover, the embedding provided by the operator $\vecd(\cdot)$ is isometric (see \eqref{eq:Frobenius} in Appendix \ref{sec:vecd}) and so the Euclidean mean can be thought of as the extrinsic mean according to the Euclidean distance in $\R^q$, $q=p(p+1)/2$.

If the measure $Q$ is defined on a Riemannian manifold with geodesic distance $\rho$, then the Fr\'{e}chet mean associated with this distance is called the intrinsic mean, not requiring an embedding. In the case of $\Sym^+(p)$, the canonical geometric mean of a measure $Q$ is defined as the intrinsic mean of $Q$ according to the canonical geodesic distance defined via
\begin{equation}
\label{eq:canonical-distance}
d_{\rm C}^2(P,X) = |\log (M^{-1/2} X M^{-1/2})|^2 = \tr\left\{[\log (M^{-1/2} X M^{-1/2})]^2\right\},
\end{equation}
where the norm is again the Frobenius norm. With this metric, $\Sym^+(p)$ becomes a geodesically complete Riemannian manifold (see Appendix \ref{sec:canonical-geom}). The canonical geometric mean is characterized by the following result.

\begin{prop}
\label{prop:canonical-geometric-mean}
Let $Q$ be a probability measure on $\Sym^+(p)$ such that $F_{\rm C}(P) = \int d_{\rm C}^2(P,X) \,Q(dX)$ is finite for $P \in \Sym^+(p)$, where $d_{\rm C}$ is the canonical geodesic distance \eqref{eq:canonical-distance}. Then the canonical geometric mean $M$ is the unique solution to the equation
\begin{equation}
\label{eq:canonical-geometric-mean}
\int \log\big(M^{-1/2} X M^{-1/2}\big) \,Q(dX) = 0.
\end{equation}
\end{prop}

This result is a direct consequence of Theorem 2.1 of \citet{Bhattacharya:2003} for complete simply connected Riemannian manifolds. The uniqueness is guaranteed by the non-positive curvature of $\Sym^+(p)$ with the canonical metric (Appendix \ref{sec:canonical-geom}). Equation \eqref{eq:canonical-geometric-mean} indicates the desired property that the residuals about the intrinsic mean, defined here as the image of the data points onto the tangent space at $M$ via the Riemannian logarithmic map (given by \eqref{eq:Log} in Appendix \ref{sec:canonical-geom}) average to zero.

The canonical geometric mean has the following equivariance properties. If $X$ is a random PD matrix according to a probability measure $Q$ with canonical geometric mean $M$, then:
\begin{enumerate}
\item The canonical geometric mean of $A X A^{-1}$ is $A M A^{-1}$, for $A \in GL(p)$.
\item The canonical geometric mean of $G X G'$ is $G M G'$, for $G \in GL(p)$.
\item The canonical geometric mean of $X^{-1}$ is $M^{-1}$.
\end{enumerate}
These properties parallel the invariance properties of the canonical distance (see \eqref{eq:canonical-distance-inv} in Appendix \ref{sec:canonical-geom}). Note the additional equivariance with respect to the group action of $GL(p)$ (see \eqref{eq:group-action} in Appendix \ref{sec:canonical-geom}) not enjoyed by the log-Euclidean mean (Section \ref{sec:log-Euclidean-avg}).

%-------------------------------------------------------------------
\subsection{The canonical geometric average in $\Sym^+(p)$}
\label{sec:canonical-avg}

For a set of i.i.d. data points $X_1,\ldots,X_n \in \Sym^+(p)$, their canonical geometric average, denoted here $\ring{X}$, is defined as their sample intrinsic mean according to the geodesic distance \eqref{eq:canonical-distance}. Applying Proposition \ref{prop:canonical-geometric-mean} to the empirical measure $\hat{Q}_n = (1/n) \sum_{i=1}^n \delta(X_i)$ gives that the canonical geometric average is unique and satisfies
\begin{equation}
\label{eq:canonical-geometric-avg}
\frac{1}{n} \sum_{i=1}^n \log\left(\ring{X}^{-1/2} X_i \ring{X}^{-1/2}\right) = 0.
\end{equation}
This condition has been also derived independently by \citet{Moakher:2005}.

Unlike the Euclidean and log-Euclidean averages, the canonical geometric average in general has no closed analytical form, except for a few cases detailed in \citet{Moakher:2005}. The following iterative algorithm is commonly used for numerically computing the canonical geometric average \citep{Pennec:1999,Fletcher:2007}.

\begin{alg}
\label{alg:intrinsic-mean}
\hfill\par\noindent
\begin{enumerate}
\item For $k=0$, set an initial value $\ring{X}^{(0)}$ such as the Euclidean average $\bar{X}$ or the log-Euclidean average $\wideparen{X}$. Fix $\epsilon > 0$.
\item Evaluate
\begin{equation*}
\bar{Y}^{(k)} = \frac{1}{n} \sum_{i=1}^n \log\left[\left(\ring{X}^{(k)}\right)^{-1/2} X_i \left(\ring{X}^{(k)}\right)^{-1/2}\right],
\qquad
\ring{X}^{(k+1)} = \left(\ring{X}^{(k)}\right)^{1/2} \exp\left(\bar{Y}^{(k)}\right) \left(\ring{X}^{(k)}\right)^{1/2}.
\end{equation*}
\item If $\left|\bar{Y}^{(k)}\right| < \epsilon$, stop. Otherwise, increase $k$ by 1 and go back to Step 2.
\end{enumerate}
\end{alg}

In Algorithm \ref{alg:intrinsic-mean}, each iteration in Step 2 maps the data onto the tangent space at $\ring{X}^{(k)}$ via the Riemannian log map, finds the Euclidean average $\bar{Y}^{(k)}$ there, and then maps it back onto the manifold via the Riemannian exponential map to obtain $\ring{X}^{(k+1)}$. The stopping condition in Step 3 reflects the property \eqref{eq:canonical-geometric-avg} that the average of the residuals should converge to zero.

Because the canonical geometric average is a special case of the canonical geometric mean with the empirical measure $\hat{Q}_n$, it has the same equivariance properties. Namely, if $X_1,\ldots,X_n$ are PD matrices with canonical geometric average $\ring{X}$, then:
\begin{enumerate}
\item The canonical geometric average of $A X_1 A^{-1}, \ldots, A X_n A^{-1}$ is $A \ring{X} A^{-1}$, for $A \in GL(p)$.
\item The canonical geometric average of $G X_1 G', \ldots, G X_n G'$ is $G \ring{X} G'$, for $G \in GL(p)$.
\item The canonical geometric average of $X_1^{-1}, \ldots, X_n^{-1}$ is $\ring{X}^{-1}$.
\end{enumerate}

The consistency of the sample canonical geometric mean is given by the following result.

\begin{prop}
\label{prop:canon-geom-avg}
Let $Q$ be a probability measure on $\Sym^+(p)$ such that $F_{\rm C}(P) = \int d_{\rm C}^2(P,X) \,Q(dX)$ is finite for $P \in \Sym^+(p)$, where $d_{\rm C}$ is the canonical geodesic distance \eqref{eq:canonical-distance}. Then the canonical geometric average $\ring{X}$ is a strongly consistent estimator of the canonical geometric mean $M$.
\end{prop}

This result is a direct consequence of Theorem 2.3 of \citet{Bhattacharya:2003} for complete metric spaces and is understood in the sense that, given any $\epsilon > 0$, the sample intrinsic mean will be within a distance less than $\epsilon$ from the true intrinsic mean for all sufficiently large sample sizes with probability 1.

%-------------------------------------------------------------------
\subsection{The canonical geometric average and the PD-matrix lognormal distribution of Type II}

Like in the case of the log-Euclidean average, the asymptotic distribution of the canonical geometric average also leads to a matrix lognormal distribution on $\Sym^+(p)$. To obtain this asymptotic distribution, we rely on the CLT for Riemannian manifolds given by Theorem 2.2 of \citet{Bhattacharya:2005}, further discussed in \citet{Bhattacharya:2008}.

\begin{thm}
\label{thm:canon-avg-CLT}
Let $X_1,\dots,X_n \in \Sym^+(p)$ be i.i.d. PD matrices according to a probability measure $Q$ that is absolutely continuous with respect to the Euclidean volume measure in $\Sym(p)$. Let $M$ and $\ring{X}$ be the canonical geometric mean and canonical geometric average, respectively. Further, define the i.i.d. random symmetric matrices
\begin{equation}
\label{eq:Y}
Y_i = \log\left(M^{-1/2} X_i M^{-1/2}\right), \quad i=1,\ldots,n
\end{equation}
and assume their common covariance matrix $\Sigma = \Cov(\vecd(Y_1))$ is finite. Then,
\begin{equation}
\label{eq:canon-avg-CLT}
\sqrt{n} \vecd\left[\log\left(M^{-1/2} \ring{X} M^{-1/2}\right)\right]
\Rightarrow N(0, K^{-1} \Sigma K^{-1})
\end{equation}
in distribution, with
\begin{equation}
\label{eq:K}
K = I + D' \E\left[\frac{1}{12}(Y_1 \ominus Y_1)^2 + \frac{1}{720}(Y_1 \ominus Y_1)^4 + \ldots\right] D,
\end{equation}
where $D$ is the $p^2\times q$ duplication matrix defined by \eqref{eq:D} in Appendix \ref{sec:vecd}, and the symbol ``$\ominus$'' denotes the operation $Y \ominus Y = Y \otimes I - I \otimes Y$.
\end{thm}

The operator ``$\ominus$'' may be called ``Kronecker difference'' in resemblance with the Kronecker sum operator $Y \oplus Y = Y \otimes I + I \otimes Y$. The matrix $K$ reflects the effect of curvature on the asymptotic covariance. The proof of Theorem \ref{thm:canon-avg-CLT} is given in Appendix \ref{sec:proofs}.

As with the arithmetic and log-Euclidean averages, the limiting distribution of the canonical geometric average is defined on the tangent space at the true value of the canonical geometric mean. Theorem \ref{thm:canon-avg-CLT} states that the canonical geometric average $\ring{X}$ is asymptotically lognormal in the sense that, for large $n$, $\log\left(M^{-1/2}\ring{X}M^{-1/2}\right)$ is approximately normally distributed in $\Sym(p)$ with mean zero and covariance $K^{-1} \Sigma K^{-1}/n$. This leads to the second definition of the lognormal distribution.

\begin{defn}
\label{defn:log-normal-distr-II}
We say that $X \in \Sym^+(p)$ has a {\em PD-matrix-variate lognormal distribution of Type II} with parameters $M \in \Sym^+(p)$ and $\Sigma \in \Sym^+(q)$, denoted $X \sim LN_{II}(M,\Sigma)$, if $\log\left(M^{-1/2} X M^{-1/2}\right) \sim N(0,\Sigma)$.
\end{defn}

Definition \ref{defn:log-normal-distr-II} has the interpretation that the parameter $M$ is precisely the canonical geometric mean of the lognormal variable $X$ because it satisfies \eqref{eq:canonical-geometric-mean}. Based on \eqref{eq:canon-avg-CLT}, Definition \ref{defn:log-normal-distr-II} allows us to write the approximate distribution of the log-Euclidean average for large $n$ as
\begin{equation}
\label{eq:canon-avg-approx}
\ring{X} \mathop{\sim}^{.} LN_{II}(M, K^{-1}\Sigma K^{-1}/n).
\end{equation}

Note that the lognormal distribution given by Definition \ref{defn:log-normal-distr-II} is not the same as the lognormal distribution given by Definition \ref{defn:log-normal-distr-I}. This is because, in general, $\log\left(M^{-1/2} X M^{-1/2}\right) \neq \log X - \log M$. The only case where equality holds for all $X \in \Sym^+(p)$ is when $M$ is a multiple of the identity matrix, i.e. $M = a I$ for some $a > 0$. In this case, the lognormal distributions of Type I and II coincide.

From Definition \ref{defn:log-normal-distr-II}, the density of $X \sim LN_{II}(M,\Sigma)$ is obtained by the change of variable $Y = \log(M^{-1/2} X M^{-1/2})$. Applying Theorem \ref{thm:log-density} to the random variable $\tilde{X} = M^{-1/2} X M^{-1/2}$, in addition to the Jacobian $J(\tilde{X} \to X) = 1/|M|$, gives that $X$ has density
\begin{equation}
\label{eq:lognormal-density-II}
f(X; M, \Sigma) = \frac{J(\tilde{X})}{(2\pi)^{q/2} |\Sigma|^{1/2} |M|}
\exp \biggl(-\frac{1}{2} \vecd(\log \tilde{X})' \Sigma^{-1}
\vecd(\log \tilde{X})\biggr)
\end{equation}
with respect to Lebesgue measure on $\Sym^+(p)$, where $J(\tilde{X})$ is given by expression \eqref{eq:Jacobian} but replacing instead the eigenvalues $\tilde{\lambda}_1 \ge \ldots \ge \tilde{\lambda}_p > 0$ of $\tilde{X} = M^{-1/2} X M^{-1/2}$, also known as the generalized eigenvalues of the pair $(M,X)$.

\begin{figure}[t]
  \centering
  a\includegraphics[height=2in,trim=0.5in 0 0.5in 0,clip]{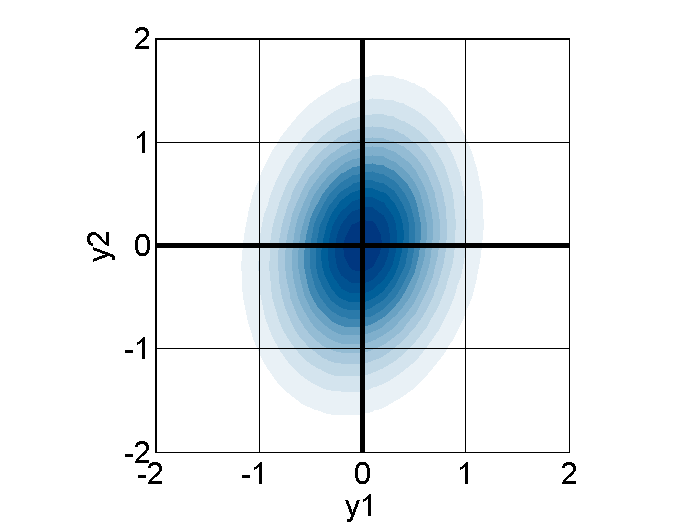}\includegraphics[height=2in,trim=0.5in 0 0.5in 0,clip]{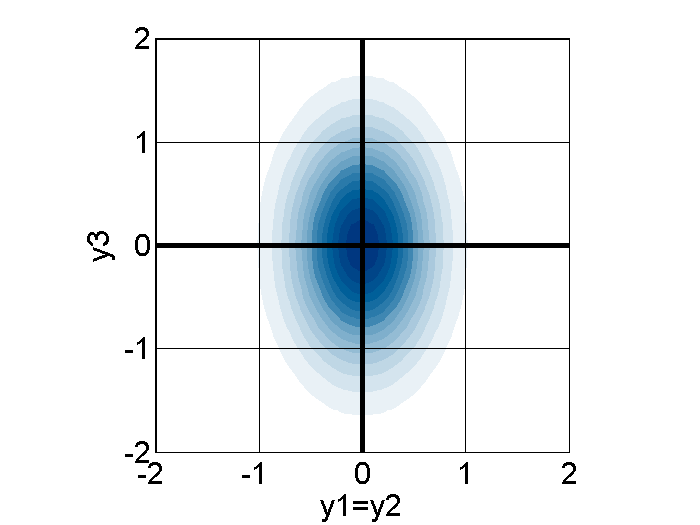}
  b\includegraphics[height=2in,trim=0.5in 0 0.5in 0,clip]{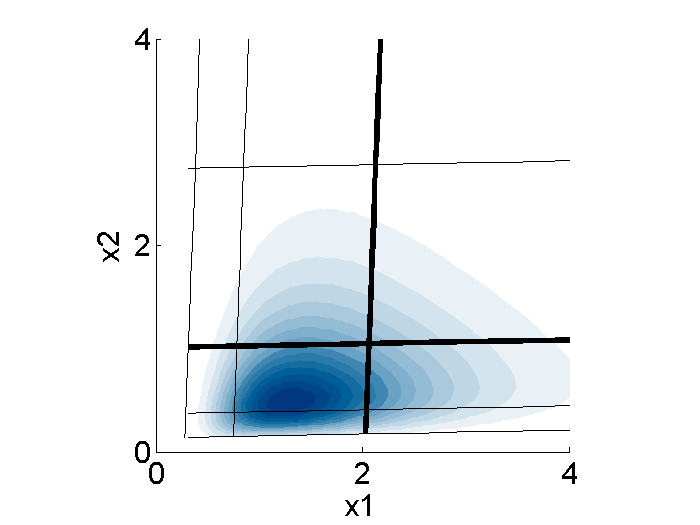}\includegraphics[height=2in,trim=0.5in 0 0.5in 0,clip]{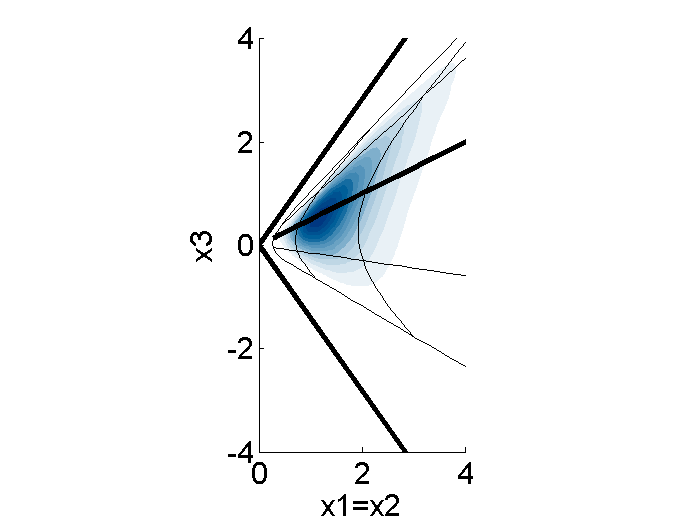}
  \caption{(a) Density of $y = \vecd(Y) \sim N(0, \Sigma)$ with parameters \eqref{eq:params-I} displayed in two planes: $y_3 = 0$ (left column) and $y_1 = y_2$ (right column). The density contours increase in height for darker colors. (b) Density of $x = \vecd(X)$ with $X = \exp(M^{1/2} Y M^{1/2})$ displayed in two planes: $x_3 = 0$ (left column) and $x_1 = x_2$ (right column).}
\label{fig:lognormal-II-density}
\end{figure}

For illustration, Figure \ref{fig:lognormal-II-density} shows the construction of a lognormal density of Type II in $\Sym^+(2)$ with the same parameters as \eqref{eq:params-I}.
\begin{comment}
\begin{equation}
\label{eq:params-II}
M = \begin{pmatrix} 2 & 0 \\ 0 & 1 \end{pmatrix}, \qquad
\Sigma = \begin{pmatrix} 0.5 & 0.25 & 0 \\ 0.25 & 1 & 0 \\ 0 & 0 & 1 \end{pmatrix}.
\end{equation}
\end{comment}
Panel (a) shows two views of the density of the normal vector $y = \vecd(Y) \sim N(0, \Sigma)$. Applying the transformation $X = \exp(M^{1/2} Y M^{1/2})$ yields the density of the vector $x = \vecd(X)$, shown in two views in Panel (b). The gridlines in Panel (a) have been mapped directly to Panel (b) in order to illustrate the deformative effect of the exponential transformation, which in contrast to Figure \ref{fig:lognormal-I-density}, is afected by $M$. Notice the shape of the $\Sym^+(2)$ cone on the right column and the additional deformation in comparison with Figure \ref{fig:lognormal-I-density}.

%-------------------------------------------------------------------
\subsection{A confidence region for the canonical geometric mean}
\label{sec:canonical-CR}

Based on \eqref{eq:canon-avg-CLT}, a confidence region for the canonical geometric mean of asymptotic level $\alpha$ is given by
\begin{equation}
\label{eq:canon-avg-CR}
\CR_{n,\alpha}(\ring{X}) = \left\{M \in \Sym^+(p): ~ n
\vecd(\ring{Y})' \hat{K} \hat{\Sigma}^{-1} \hat{K} \vecd(\ring{Y})
\le \chi^2_{q,1-\alpha}\right\}
\end{equation}
where
\[
\ring{Y} = \log\left(M^{-1/2} \ring{X} M^{-1/2}\right) %= \log\left(\ring{X}^{-1/2} M \ring{X}^{-1/2}\right).
\]
The matrices $\hat{\Sigma}$ and $\hat{K}$ in \eqref{eq:canon-avg-CR} are consistent estimators of $\Sigma$ and $K$, respectively, and may be obtained as follows. Letting
\[
\hat{Y}_i = \log\left(\ring{X}^{-1/2} X_i \ring{X}^{-1/2}\right), \quad i=1,\ldots,n
\]
in accordance to \eqref{eq:Y} where $\ring{X}$ estimates $M$, we can take
\[
\hat{\Sigma} = \frac{1}{n}\sum_{i=1}^n \vecd(\hat{Y}_i) \vecd(\hat{Y}_i)'.
\]
Note that there is no need to subtract the average of the $\vecd(\hat{Y}_i)$'s in the computation of the sample covariance because it is zero by \eqref{eq:canonical-geometric-avg}. Similarly, from \eqref{eq:K}, we can take
\[
\hat{K} = I + D' \frac{1}{n}\sum_{i=1}^n \left[\frac{1}{12}(\hat{Y}_i \ominus \hat{Y}_i)^2 + \frac{1}{720}(\hat{Y}_i \ominus \hat{Y}_i)^4 + \ldots\right] D.
\]

It is worth noting that in \eqref{eq:canon-avg-CR}, $\ring{Y}$ depends on $M$ in a nonlinear way. In the data analysis Section \ref{sec:dataExample} below we will compute the extreme points along the first principal component of the CR (in the log domain). These are defined as the points $M$ satisfying the equation
\begin{equation}
\label{eq:quadratic}
M^{-1/2} \ring{X} M^{-1/2} = B_{\pm}
\end{equation}
where
\[
B_{\pm} = \exp\left[\pm \vecd^{-1}\left( \lambda_1 V_1 \chi^2_{q,1-\alpha}/n \right)\right]
\]
and $\lambda_1$, $V_1$ are the first eigenvalue and first eigenvector of the matrix $\hat{K}^{-1} \hat{\Sigma} \hat{K}^{-1}$.
Equation \eqref{eq:quadratic} is a special case of the continuous-time algebraic Ricatti equation. Interestingly, its solution is also a canonical geometric average in the following sense. By \citet{Lawson:2001}, the unique solution to the equation $X A^{-1} X = B$ for $A, B \in \Sym^+(p)$ is the canonical geometric average of $A$ and $B$ defined via \eqref{eq:canonical-geometric-avg} which, for $n=2$, has the closed form solution $A\# B = A (A^{-1} B)^{1/2}$ \citep{Moakher:2005}. Therefore the solution to \eqref{eq:quadratic} is
\[
M = \left[\left(\ring{X} B\right)^{-1/2} \ring{X}\right]^2.
\]

%===================================================================
\section{A DTI data example}
\label{sec:dataExample}

%-------------------------------------------------------------------
\subsection{Data description}
\label{sec:data}

Diffusion Tensor Imaging (DTI) is a particular modality of magnetic resonance imaging often used to visualize the brain's white matter \citep{Basser:1996,LeBihan:2001}. DTI images are 3D rectangular arrays that contain at every voxel a $3 \times 3$ symmetric PD matrix, also called diffusion tensor (DT). %The DT can be interpreted as the covariance matrix of a 3D Gaussian distribution that models the Brownian motion of the water molecules in the voxel, and is reconstructed from measurements of the diffusion coefficient in at least 6 directions in space.
A useful scalar summary of the DT is fractional anisotropy (FA), a function of the DT's eigenvalues. FA values near 0 represent nearly isotropic diffusion, found in the brain's water ventricles, while values near 1 represent highly anisotropic diffusion, found in tighly packed white matter neural fibers. Another useful summary is the principal diffusion direction (PDD), the first eigenvector of the DT. In highly anisotropic voxels it indicates the direction of the underlying neural fibers.

In this paper, we analyze a dataset consisting of 34 DTI images corresponding to 34 10-year-old children (12 boys and 22 girls). This dataset was also analyzed in \citet{Schwartzman:2010a} and is part of a larger observational study of brain anatomy in children \citep{Dougherty:2007}. The 34 DTI images were spatially normalized to a common template \citep{Dougherty:2005}, resulting in each image being a voxel array of size $81 \times 106 \times 76$ in a rectangular grid with $2 \times 2 \times 2$ mm regular spacings. The coordinate axes are defined so that the $x$, $y$ and $z$ axis point respectively to the right, front and top of the brain (Figure \ref{fig:DT-averages}). Like in \citet{Schwartzman:2010a}, analysis was restricted to voxels strictly inside the brain.

%-------------------------------------------------------------------
\subsection{Data analysis}
\label{sec:dataAnalysis}

After spatial normalization, the collection of 34 DT's across subjects at each voxel can be modeled as an i.i.d. sample of PD matrices whose population mean and covariance depends on the voxel location. As such, estimation of the population mean DT image is achieved by computing PD averages voxelwise \citep{Schwartzman:2010a}.

\begin{figure}
  \centering
  \begin{tabular}{ccc}
  \includegraphics[height=2in]{\figurepath/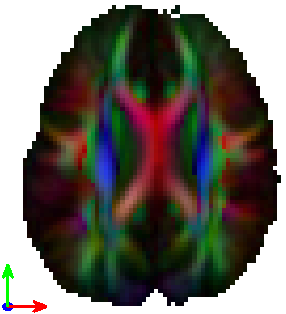} &
  \includegraphics[height=2in]{\figurepath/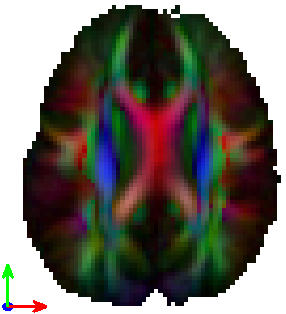} &
  \includegraphics[height=2in]{\figurepath/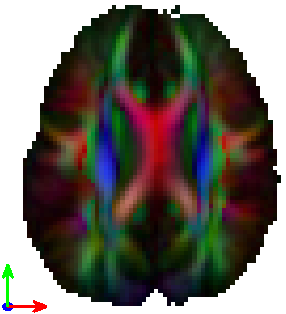} \\
  Euclidean Avg. & Log-Euclidean Avg. & Canonical Geom. Avg.
  \end{tabular}
  \caption{Voxelwise DT averages. Image intensity is proportional to FA in the range $[0,0.6]$. Colors indicate coordinate directions of the PDD: right-left (red), anterior-posterior (green) and superior-inferior (blue). Mixed colors represent directions that are oblique to the coordinate axes.}
  \label{fig:DT-averages}
\end{figure}

Figure \ref{fig:DT-averages} shows the three types of voxelwise averages discussed in this paper at a transverse slice 36 mm above the anterior commisure, an anatomical landmark commonly used for spatial normalization (Talairach convention). This is the same slice that is shown in \citet{Schwartzman:2010a}. Because DTs are 6-dimensional objects, Figure \ref{fig:DT-averages} shows a meaningful reduction to 3 dimensions. One dimension is captured by FA, encoded in the image intensity, while the other two are captured by the PDD, encoded in color.

\begin{figure}[t!]
  \centering
  \begin{tabular}{rccl}
  a & \includegraphics[height=2in]{\figurepath/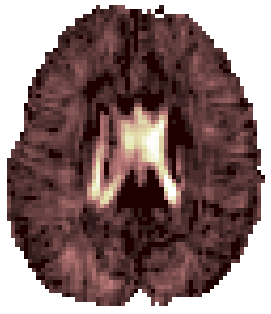}
    & \includegraphics[height=2in]{\figurepath/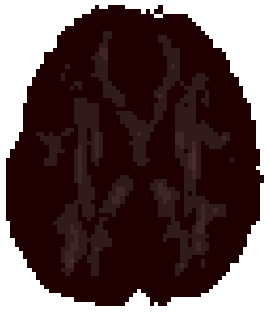}
    & \includegraphics[height=1.1in]{\figurepath/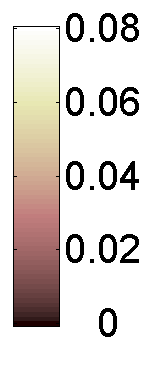} \\
  b & \includegraphics[height=2in]{\figurepath/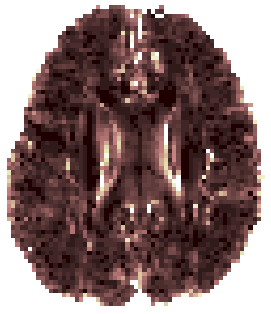}
    & \includegraphics[height=2in]{\figurepath/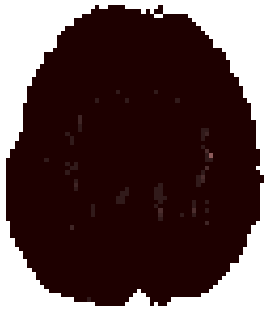}
    & \includegraphics[height=1.1in]{\figurepath/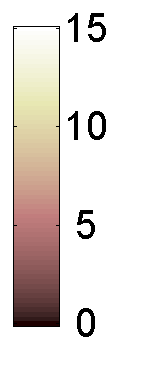}
  \end{tabular}
  \caption{(a) FA differences between: log-Euclidean and Euclidean average (left); log-Euclidean and canonical (right). Image intensity scale is the same for both panels. (b) Angle between: PDD of log-Euclidean and Euclidean average (left); PDD of log-Euclidean and canonical (right). Image intensity scale in degrees is the same in both panels.}
%  \caption{Differences between voxelwise DT averages: (a) FA of log-Euclidean average minus FA of Euclidean average (left); FA of log-Euclidean average minus FA of canonical geometric average (right). Image intensity scale is the same for both panels. (b) Angle beween PDD of log-Euclidean average and PDD of Euclidean average (left); Angle beween PDD of log-Euclidean average and PDD of canonical geometric average (right). Image intensity scale in degrees is the same in both panels.}
  \label{fig:Avg-diff}
\end{figure}

At first, all three averages appear to be very similar. However, Figure \ref{fig:Avg-diff} shows that the Euclidean and log-Euclidean averages disagree in some regions, particularly in the corpus callosum (red X-like structure in Figure \ref{fig:DT-averages}). The differences between the log-Euclidean and canonical geometric averages are about an order of magnitude smaller.
%taking differences between FA values, as shown in Figure \ref{fig:Avg-diff}a, reveals that the Euclidean and log-Euclidean averages disagree in some regions, particularly in the corpus callosum (red X-like structure in Figure \ref{fig:DT-averages}), while the differences between the log-Euclidean and canonical geometric averages are about an order of magnitude smaller.
%Similarly, the Euclidean and log-Euclidean averages exhibit large differences in the orientation of the PDD, as shown in Figure \ref{fig:Avg-diff}b, with angles in some voxels as large as $70^\circ$ or $80^\circ$. On the other hand, the difference in orientation between the log-Euclidean and canonical geometric averages do not exceed $2^\circ$.

In the current debate about average types, the DTI literature has not yet considered whether the observed differences between them may be due to sampling variability. Constructing a formal two-sample test for a pair of means is difficult because they are defined by different geometries and because of the dependence between them, as they are based on the same data. More informally, to make use of the CRs developed above, we may check whether the value of a particular type of average is inside the CR of another type of average. This approach is conservative in detecting differences because the variability in any type of average is smaller than the variability in the difference between two types of averages.

\begin{figure}
  \centering
  \begin{tabular}{cccl}
  a \includegraphics[height=2.0in]{\figurepath/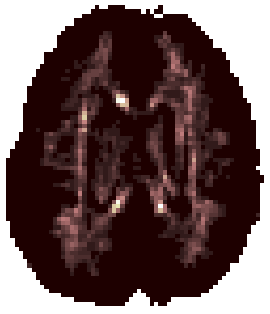}
  & b \includegraphics[height=2.0in]{\figurepath/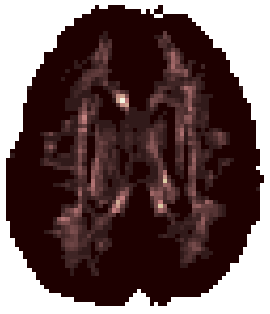}
  & c \includegraphics[height=2.0in]{\figurepath/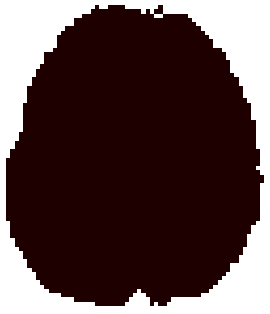}
  & \includegraphics[height=1.1in]{\figurepath/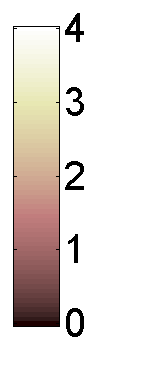}
  \end{tabular}
  \caption{$\Log_{10}$ p-values of voxelwise DT averages: (a) Log-Euclidean average with respect to Euclidean CR; (b) Euclidean average with respect to log-Euclidean CR; (c) Canonical geometric average with respect to log-Euclidean CR. Image intensity scale is the same in all panels.}
  \label{fig:pImg}
\end{figure}

Following this approach, Figure \ref{fig:pImg}a evaluates the log-Euclidean average with respect to the Euclidean average. Shown is a map of the smallest $\alpha$ such that the log-Euclidean average $\wideparen{X}$ is inside the Euclidean CR defined by \eqref{eq:Euclidean-avg-CR}. This may be loosely interpreted as a p-value map for the observed log-Euclidean average under the assumption that the Euclidean average is the true mean. Interestingly, only 28 out of the 3419 voxels in this slice have a p-value smaller than 0.05, and only 4 of them have a p-value smaller than the $2.3\times 10^{-4}$ threshold corresponding to an FDR of 0.2 using the Benjamini-Hochberg algorithm \citep{Benjamini:1995}.

To double-check, Figure \ref{fig:pImg}b does the reverse and evaluates the Euclidean average with respect to the log-Euclidean average by showing a map of the smallest $\alpha$ such that the Euclidean average $\bar{X}$ is inside the log-Euclidean CR defined by \eqref{eq:log-Euclidean-avg-CR}. This may be interpreted loosely as a p-value map for the observed Euclidean average under the assumption that the log-Euclidean average is the true mean. In this comparison, only 16 out of the 3419 voxels in this slice have a p-value smaller than 0.05, and none of them survive the 0.2 FDR cutoff. In summary, most of the differences between the Euclidean and log-Euclidean averages are indeed within the range of random variation.

For completeness, Figure \ref{fig:pImg}c shows the corresponding p-value map comparing the canonical geometric average to the log-Euclidean CR. Here all p-values are very close to 1. The results are similar when comparing the log-Euclidean average to the canonical geometric CR (map not shown). In other words, the log-Euclidean and canonical geometric averages are statistically indistinguishable.

\begin{figure}
  \centering
  \begin{tabular}{ccc}
  \includegraphics[height=1.7in]{\figurepath/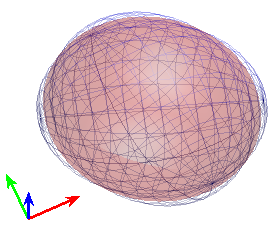} &
  \includegraphics[height=1.7in]{\figurepath/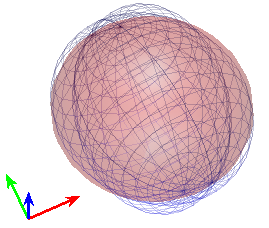} &
  \includegraphics[height=1.7in]{\figurepath/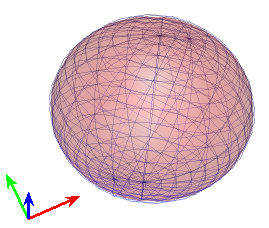} \\
  Euclidean Avg. & Log-Euclidean Avg. & Canonical Geom. Avg.
  \end{tabular}
  \caption{The voxel with the largest differences between the average types (brightest voxel in Figure \ref{fig:pImg}a): ellipsoidal rendering of the DT averages (red) and two extreme points of the corresponding 95\% CR (blue). Colored axes correspond to the coordinate directions described in Figure \ref{fig:DT-averages}.}
  \label{fig:DT-ellipsoids}
\end{figure}

To better visualize the few significant differences between the three types of averages, Figure \ref{fig:DT-ellipsoids} shows their ellipsoidal rendering via \eqref{eq:ellipsoid} for the voxel with the smallest p-value in Figure \ref{fig:pImg}a. Superimposed are similar ellipsoidal renderings of the extreme points along the first principal component (in the log domain) of the corresponding 95\% CR, as was done for the 2D example in Figure \ref{fig:random-ellipses}. The observed averages do exhibit visible differences in FA and orientation, but so do the extreme points of the CRs. Note that these 6-dimensional extreme points do not show the full extent of the 21-dimensional CRs but only their main mode of variation.

%===================================================================
\section{Discussion}
\label{sec:discussion}

%-------------------------------------------------------------------
\subsection{Metrics and averages in $\Sym^+(p)$}

In this paper we have considered the three most common types of averages and their corresponding population means defined on $\Sym^+(p)$: arithmetic or Euclidean, log-Euclidean and canonical geometric. We have seen that the CLT applied to the latter two leads to the definition of a lognormal distribution on $\Sym^+(p)$, analogous to the lognormal distribution for positive scalars.

These three types of averages correspond to three different metrics defined on $\Sym^+(p)$. The preference for a particular metric and corresponding average type may is often based on qualitative interpretation of the results in DTI analysis \citep{Fletcher:2007,Whitcher:2007,Dryden:2009,Pasternak:2010}. Often for data on manifolds, such as data on the sphere, Riemannian metrics are appealing because they lead to intrinsic means instead of extrinsic means. Computational considerations are also important \citep{Groisser:2004,Patrangenaru:2012}. However, in the case of $\Sym^+(p)$ with the Euclidean metric, the intrinsic and extrinsic means coincide and are equal to the simple Euclidean average, which has a well defined asymptotic normal distribution despite $\Sym^+(p)$ not being a vector space. From this point of view, going beyond Euclidean metrics for $\Sym^+(p)$ may seem unnecessary.

In this paper, we have considered the random variability of each average. According to the data analysis of Section \ref{sec:dataExample}, the log-Euclidean and canonical geometric averages are virtually indistinguishable, while the Euclidean and log-Euclidean averages are mostly within the expected random variation for the given sample size. When comparing their computational complexity, our Matlab implementation computed the Euclidean, log-Euclidean and canonical geometric averages for all 3419 voxels in the slice shown in 0.72 s, 3.48 s and 16.48 s, respectively. These results make the log-Euclidean average convincingly preferable to the canonical geometric average. The choice between the Euclidean and log-Euclidean averages may be based on their desired equivariance properties, but perhaps is only important for larger sample sizes. The statistical relevance of other types of averages suggested by \citet{Dryden:2009} could be analyzed using large sample techniques like the ones used in this paper.

%-------------------------------------------------------------------
\subsection{Parametric models for random PD matrices}

The two PD-matrix-variate lognormal families of distributions defined in this paper were obtained by applying the CLT to the log-Euclidean and canonical geometric averages of PD matrices. We derived their explicit densities and provided CRs for the corresponding geometric means.

More generally, the PD-matrix-variate lognormal families could be used as parametric models for PD matrix data, either for frequentist data analysis or as flexible priors for Bayesian analysis. The lognormal distributions provide a more flexible parametric model than the Wishart distribution because it has $q = p(p+1)/2$ parameters for the mean and $q(q+1)/2$ separate additional parameters for the covariance, while the Wishart distribution has $q$ parameters for the mean and one parameter for the number of degrees of freedom, which affects both the covariance and the mean. The matrix lognormal distribution, on the other hand, is heavier-tailed than the Wishart, just like the scalar lognormal is heavier-tailed than the $\chi^2$ distribution.

Going beyond PD matrices, the concept of a matrix-variate lognormal distribution could be extended to other Riemannian manifolds where the Riemannian logarithmic map is the matrix logarithm. One example is the orthogonal group $O(p)$, whose tangent space is the set of antisymmetric or skew-symmetric matrices. Extrinsic and intrinsic averages on that set have been studied by \citet{Moakher:2002}. Applying the CLT to the log-Euclidean average of orthogonal matrices should lead naturally to a lognormal distribution on that set. The details are left for future work.

%===================================================================
\section*{Acknowledgments}
Special thanks to Emmanuel Ben-David (Columbia U.), Victor Patrangenaru (Florida State U.) and David Groisser (U. of Florida, Gainsville) for helpful discussions, as well as Bob Dougherty (Stanford U.) for providing the DTI data. This work was partially funded by NIH grant R21EB012177.
% syasamin@indiana.edu

%===================================================================
\appendix

\section{Auxiliary definitions and results}
\label{sec:basics}

%-------------------------------------------------------------------
\subsection{The $\vecd(\cdot)$ operator}
\label{sec:vecd}

For a matrix $Y \in \Sym(p)$, we define the mapping of $\Sym(p)$ to $\R^q$ with $q=p(p+1)/2$ given by
\begin{equation}
\label{eq:vecd}
\vecd(Y) = (\diag(Y)', \sqrt{2}\,\offdiag(Y)')',
\end{equation}
where $\diag(Y)$ is a $p \times 1$ column vector containing the diagonal entries of $Y$ and $\offdiag(Y)$ is a $(q-p) \times 1$ column vector containing the off-diagonal entries of $Y$ copied from below the diagonal columnwise (or above the diagonal rowwise) \citep{Schwartzman:2010a}. For example, in the case $p=3$, $q=6$,
$$
\vecd(Y) = (Y_{11},Y_{22},Y_{33},\sqrt{2}\,Y_{12},\sqrt{2}\,Y_{13},\sqrt{2}\,Y_{23})'.
$$

In contrast to the usual columnwise vectorization operator $\vec(\cdot)$ \citep{Gupta:2000,Chiu:1996}, $\vecd(\cdot)$ provides a more natural representation for data in the form of symmetric matrices and it has the convenient property that the Frobenius norm of $Y$ is the same as the Euclidean norm of $\vecd(Y)$, i.e.
\begin{equation}
\label{eq:Frobenius}
\tr(Y^2) = \vecd(Y)' \vecd(Y).
\end{equation}
The operator $\vecd(\cdot)$ provides an embedding of both $\Sym(p)$ and $\Sym^+(p)$ in $\R^q$. This is the same embedding illustrated for $p=2$ in Figure \ref{fig:cone}. The operator $\vecd(\cdot)$ is invertible and its inverse shall be denoted by $\vecd^{-1}(\cdot)$.

For any $p$, a duplication matrix $D$ of size $p^2 \times q$ exists such that the operator $\vecd(\cdot)$ is related to the usual columnwise vectorization operator $\vec(\cdot)$ by
\begin{equation}
\label{eq:D}
\vec(Y) = D \vecd(Y), \qquad \vecd(Y) = D' \vec(Y).
\end{equation}
For example, in the case $p=3$, $q=6$,
\[
D = \begin{pmatrix}
1 & 0 & 0 & 0          & 0          & 0 \\
0 & 0 & 0 & 1/\sqrt{2} & 0          & 0 \\
0 & 0 & 0 & 0          & 1/\sqrt{2} & 0 \\
0 & 0 & 0 & 1/\sqrt{2} & 0          & 0 \\
0 & 1 & 0 & 0          & 0          & 0 \\
0 & 0 & 0 & 0          & 0          & 1/\sqrt{2} \\
0 & 0 & 0 & 0          & 1/\sqrt{2} & 0 \\
0 & 0 & 0 & 0          & 0          & 1/\sqrt{2} \\
0 & 0 & 1 & 0          & 0          & 0
\end{pmatrix}.
\]
The duplication matrix $D$ has rank $q$ and it satisfies the following properties. For any $Y, A \in \Sym(p)$,
%\begin{subequations}
\begin{eqnarray*}
D'D & = & I \\
D D' \vec(Y) & = & \vec(Y) \\
\vecd(A Y A') & = & D' (A \otimes A) D \vecd(Y),
\end{eqnarray*}
%\end{subequations}
where the third property is based on the general property that $\vec(A Y B') = (B \otimes A) \vec(Y)$ \citep{Schott:2005}.

%-------------------------------------------------------------------
\subsection{Matrix exponential and logarithm}
\label{sec:exp-log}

The exponential of a square matrix $Y$ is defined as $X = \exp(Y) = e^Y = \sum_{k=0}^\infty Y^k/k!$. The inverse operation $Y = \log(X)$ is called the matrix logarithm of $X$. In particular, if $Y \in \Sym(p)$ and $Y = V L V'$ is an eigen-decomposition with $V \in O(p)$ and $L \in \Diag(p)$, then it follows from the series expansion that
\begin{equation}
\label{eq:exp}
X = \exp(Y) = V \exp(L) V',
\end{equation}
where $\exp(L)$ is a diagonal matrix containing the exponential of each of the diagonal elements of $L$. Since the diagonal elements of $L$ are real, the diagonal elements of $\exp(L)$ are positive, implying that $X$ is PD. Given $X \in \Sym^+(p)$, \eqref{eq:exp} can be inverted and the matrix log of $X$ is uniquely given by
\begin{equation*}
\label{eq:log}
Y = \log(X) = V \log(\Lambda) V',
\end{equation*}
where $\log(\Lambda)$ is a diagonal matrix containing the logs of each of the diagonal elements of $\Lambda$.

%This relationship establishes a one-to-one correspondence between $\Sym^+(p)$ and $\Sym(p)$, as in \eqref{eq:diffeomorphism}. Figure \ref{fig:cone} shows that this correspondence allows extrapolation between elements of $\Sym^+(p)$ in such a way that the result stays within $\Sym^+(p)$, which is not guaranteed if $\Sym^+(p)$ is treated as merely a subset of $\Sym(p)$.

Both $\exp(\cdot)$ and $\log(\cdot)$ are analytical functions and enjoy the useful properties with respect to similarity transformations and matrix inversion
\begin{equation}
\label{eq:analytical}
\begin{alignedat}{2}
\exp\big(A Y A^{-1}\big) &= A \exp(Y) A^{-1}, &\qquad \big(e^Y\big)^{-1} &= e^{-Y} \\
\log\big(A X A^{-1}\big) &= A \log(X) A^{-1}, &\qquad \log\big(X^{-1}\big) &= -\log(X)
\end{alignedat}
\end{equation}
for $A \in GL(p)$, the set of $p \times p$ real invertible matrices. In general, $e^{Y + Z} \ne e^{Y}e^{Z}$ and $\log(XW) \ne \log(X) + \log(W)$ unless $Y$ and $Z$ (or $X$ and $W$) commute, i.e. they have the same eigenvectors. Instead, there is the Baker-Campbell-Hausdorff (BCH) formula
\begin{equation*}
\label{eq:BCH}
\log(e^Y e^Z) = Y + Z + \frac{1}{2}[Y,Z]
+ \frac{1}{12}[Y,[Y,Z]] + \frac{1}{12}[Z,[Z,Y]] + \frac{1}{24}[[Y,[Y,Z]],Z] + \ldots\end{equation*}
where $[Y,Z] = YZ - ZY$ is the matrix commutator. Notice that indeed $\log(e^Y e^Z) = Y + Z$ if $Y$ and $Z$ commute. In this paper we use an explicit polynomial expression without the commutators, also known as Goldberg's expansion \citep{Goldberg:1956}. Using the method of \citet{Newman:1987}, the first several terms of this series are found to be
\begin{equation}
\label{eq:Goldberg}
\begin{aligned}
\log(e^Y & e^Z) = Y + Z + \frac{1}{2}(YZ - ZY) + \frac{1}{12}\big(Y Z^2 + Y^2 Z + Z Y^2 + Z^2 Y) - \frac{1}{6}\big(Y Z Y + Z Y Z\big) \\
&+ \frac{1}{24}\big(Y^2 Z^2 - Z^2 Y^2\big) - \frac{1}{12}\big(Y Z Y Z - Z Y Z Y\big) - \frac{1}{720}\big(Y Z^4 + Y^4 Z + Z Y^4 + Z^4 Y\big) \\
&+ \frac{1}{180}\big(Y^2 Z^3 + Y^3 Z^2 + Z^2 Y^3 + Z^3 Y^2
+ Y Z Y^3 + Y Z^3 Y + Y^3 Z Y + Z Y Z^3 + Z Y^3 Z + Z^3 Y Z\big) \\
&- \frac{1}{120}\big(Y Z^2 Y^2 + Y^2 Z Y^2 + Y^2 Z^2 Y + Z Y^2 Z^2 + Z^2 Y Z^2 + Z^2 Y^2 Z\big) \\
&- \frac{1}{120}\big(Y Z Y Z^2 + Y Z Y^2 Z + Y Z^2 Y Z + Y^2 Z Y Z + Z Y Z Y^2 + Z Y Z^2 Y + Z Y^2 Z Y + Z^2 Y Z Y\big) \\
&+ \frac{1}{30}\big(Y Z Y Z Y + Z Y Z Y Z\big) + \ldots
\end{aligned}
\end{equation}

%-------------------------------------------------------------------
\subsection{$\Sym^+(p)$ as a Riemannian manifold with the Euclidean and log-Euclidean metrics}
\label{sec:Euclidean-geom}

In the view of Section \ref{sec:Euclidean}, $\Sym^+(p)$ is seen as a Riemannian submanifold of $\Sym(p)$ endowed with the Euclidean or Frobenius metric inherited from $\Sym(p)$. Given two tangent vectors $Y,Z \in \Sym(p)$, their Euclidean or Frobenius inner product and associated squared norm are defined as
\begin{equation}
\label{eq:Frobenius-inner-prod}
\langle Y, Z \rangle = \tr(Y Z) = \vecd(Y)' \vecd(Z),
\qquad |Y|^2 = \tr(Y^2),
\end{equation}
and the geodesic distance between two points $X_1,X_2 \in \Sym^+(p)$ is the Euclidean distance defined via \eqref{eq:Euclidean-distance}.
%\begin{equation}
%\label{eq:Euclidean-distance}
%d_{\rm E}(X_1,X_2) = |X_1 - X_2| = \left[\tr(X_1 - X_2)^2\right]^{1/2}.
%\end{equation}

A Riemannian manifold is called geodesically complete if at any point on the manifold, the Riemannian exponential map, which maps tangent vectors to the manifold along geodesics, is defined for all tangent vectors. This property depends on the metric that is defined on the manifold. When endowed with the Euclidean metric, $\Sym^+(p)$ is not geodesically complete because the Riemannian exponential map from the tangent space $\Sym(p)$ to $\Sym^+(p)$ at any point, which for the Euclidean metric is equal to a displacement by a tangent vector, is only defined for those tangent vectors whose displacement falls within the boundaries of $\Sym^+(p)$. As a consequence, the approximate normal distribution of the Euclidean average \eqref{eq:Euclidean-avg-approx} is defined on $\Sym(p)$ but not on $\Sym^+(p)$.

One way of achieving geodesic completeness is via the global matrix log transformation \eqref{eq:diffeomorphism}, which effectively removes the PD constraints by mapping each element of $\Sym^+(p)$ to a unique element of $\Sym(p)$ \citep{Arsigny:2006,Schwartzman:Thesis2006}. This is the view of Sections \ref{sec:log-Euclidean-avg} and \ref{sec:log-Euclidean-distr-I}. According to this metric, the geodesic distance between two points $X_1,X_2 \in \Sym^+(p)$ is the log-Euclidean distance defined via \eqref{eq:log-Euclidean-distance}.
%\begin{equation*}
%\label{eq:log-Euclidean-distance}
%d_{\rm LE}(X_1,X_2) = |\log X_1 - \log X_2| = \left[\tr(\log X_1 - \log X_2)^2\right]^{1/2}.
%end{equation*}
%When endowed with the log-Euclidean metric, $\Sym^+(p)$ is geodesically complete. As a consequence,
Because of geodesic completeness, the approximate lognormal distribution of the log-Euclidean average \eqref{eq:log-Euclidean-avg-approx} is well defined on $\Sym^+(p)$.

As shown in Sections \ref{sec:log-Euclidean-avg} and \ref{sec:log-Euclidean-distr-I}, the log transformation is simple enough that it permits obtaining results without the need of the Riemannian machinery.

%-------------------------------------------------------------------
\subsection{$\Sym^+(p)$ as a Riemannian manifold with the canonical metric}
\label{sec:canonical-geom}

A more elaborate way of producing a geodesically complete manifold is via the so-called canonical or affine-invariant metric. For tangent vectors $Y, Z \in \Sym(p)$ at $M \in \Sym^+(p)$, their canonical or affine invariant inner product and associated squared norm are defined as
\begin{equation}
\label{eq:canonical-inner-prod}
\Langle Y, Z \Rangle_M = \tr(M^{-1} Y M^{-1} Z), \qquad
\|Y\|^2_M = \tr\left[(M^{-1} Y)^2\right].
\end{equation}
Notice that at $M = I$, the $p\times p$ identity matrix, the canonical inner product \eqref{eq:canonical-inner-prod} reduces to the Frobenius inner product \eqref{eq:Frobenius-inner-prod}.
Various aspects of the geometry of $\Sym^+(p)$ under the metric \eqref{eq:canonical-inner-prod} have been described elsewhere \citep{Kobayashi:1996,Lang:1999,Moakher:2005,Smith:2005,Schwartzman:Thesis2006,Lenglet:2006,Zhu:2009}, which we summarize here for completeness and easier reference.

Define the linear group action of $GL(p)$ on $\Sym^+(p)$
\begin{equation}
\label{eq:group-action}
\phi(G, X) = G X G', \qquad G \in GL(p), \qquad X \in \Sym^+(p),
\end{equation}
which maps the PD matrix $X$ to the PD matrix $G X G'$. The metric \eqref{eq:canonical-inner-prod} is called affine-invariant because it has the isometric property of being invariant under the group action \eqref{eq:group-action}, that is
\begin{equation}
\label{eq:isometry}
\Langle Y, Z \Rangle_M = \Langle G Y G', G Z G' \Rangle_{G M G'}
\end{equation}
for any $G \in GL(p)$. In particular, if $G = M^{-1/2}$, then \eqref{eq:isometry} becomes
\[
\Langle Y, Z \Rangle_M = \langle M^{-1/2} Y M^{-1/2}, M^{-1/2} Z M^{-1/2} \rangle_I = \tr(M^{-1/2} Y M^{-1/2} \cdot M^{-1/2} Z M^{-1/2}).
\]
This says that the affine-invariant inner product can be computed by translating the tangent vectors to the tangent space at the identity and computing the standard Frobenius inner product there.

This device is useful for obtaining the Riemannian exponential and logarithmic maps corresponding to the affine-invariant metric. Since the Riemannian exponential map of a tangent vector $Y \in \Sym(p)$ at the identity $I$ is equal to the matrix exponential $\exp(Y) \in \Sym^+(p)$, the Riemannian exponential map of a tangent vector $Y$ at $M$ onto $\Sym^+(p)$ is obtained by mapping $Y$ to the tangent space at the identity via $\phi(M^{-1/2},\cdot)$, evaluating the Riemannian exponential map there, and mapping the result back again via $\phi(M^{1/2},\cdot)$. The Riemannian exponential map is thus given by
\begin{equation}
\label{eq:Exp}
X = \Exp_M(Y) = M^{1/2} \exp(M^{-1/2} Y M^{-1/2}) M^{1/2} = M \exp(M^{-1} Y),
\end{equation}
where the second expression is obtained via \eqref{eq:analytical}.
By \eqref{eq:diffeomorphism}, this map is one-to-one. The inverse map is the Riemannian logarithmic map of $X \in \Sym^+(p)$ to the tangent vector $Y \in \Sym(p)$ in the tangent space at $M$, and is given by
\begin{equation}
\label{eq:Log}
Y = \Log_M(X) = M^{1/2} \log(M^{-1/2} X M^{-1/2}) M^{1/2} = M \log(M^{-1} X).
\end{equation}

Because the Riemannian exponential map is bijective, there is a unique geodesic path joining any two points in $\Sym^+(p)$. The geodesic distance between them may be defined uniquely as the length of that path, which by Gauss' lemma \citep[p. 69]{DoCarmo:1992} is equal to the length of the projection via the Riemannian logarithmic map of one of the points onto the tangent space at the other. For any two points $M, X \in \Sym^+(p)$, their geodesic distance according to the affine-invariant metric \eqref{eq:canonical-inner-prod} is defined via \eqref{eq:canonical-distance}.
%\begin{equation}
%\label{eq:canonical-distance}
%d_{\rm C}(M,X) = \|\Log_M X\|_M = |\log (M^{-1/2} X M^{-1/2})| = |\log(M^{-1} X)|.
%\end{equation}

The bijective property of the Riemannian exponential map makes $\Sym^+(p)$ geodesically complete. Consequently, it is also complete as a metric space by the Hopf-Rinow theorem \citep[p. 147]{DoCarmo:1992}. The geodesic distance \eqref{eq:canonical-distance} is a proper metric and satisfies the properties of semi-positiveness, symmetry and the triangle inequality. In addition, it is invariant under similarity transformations, under the group action \eqref{eq:group-action} and under inversion \citep{Forstner:1999}:
\begin{equation}
\label{eq:canonical-distance-inv}
\begin{aligned}
d_{\rm C}(M,X) &= d_{\rm C}(AMA^{-1},AXA^{-1}), \qquad \forall A \in GL(p) \\
d_{\rm C}(M,X) &= d_{\rm C}(GMG',GXG'), \qquad \forall G \in GL(p) \\
d_{\rm C}(M,X) &= d_{\rm C}(M^{-1},X^{-1}).
\end{aligned}
\end{equation}
These properties make $\Sym^+(p)$ a Riemannian homogeneous space and a symmetric space in the sense of Cartan \citep[Ch. X-XI]{Kobayashi:1996}.

The curvature of $\Sym^+(p)$ under the affine-invariant inner product is non-positive and has been derived explicitly by \citet{Lenglet:2006}. Intuitively, non-positive curvature means that geodesics ``diverge'' away from each other. This property guarantees the existence and uniqueness of the canonical geometric mean (Proposition \ref{prop:canonical-geometric-mean}).

%===================================================================
\section{Proofs}
\label{sec:proofs}

\subsection{Proof of Theorem \ref{thm:log-density}}
\par
Let $X = V \Lambda V'$ and $Y = V L V'$ be eigen-decompositions of $X$ and $Y$ so that $L = \log \Lambda$ with diagonal entries $l_i = \log \lambda_i$. Suppose the eigenvalues are all distinct. The Jacobian of the eigenvalue decomposition $X \to (V,\Lambda)$ can be written as $\mathcal{J}\big(X \to (V,\Lambda)\big) = g(V) \prod_{i<j} (\lambda_i - \lambda_j)$, where $g(V)$ is a function of the eigenvectors only \citep[p. 58]{Mehta:1991}. Similarly, $\mathcal{J}\big(Y \to (V,L)\big) = g(V) \prod_{i<j} (l_i - l_j)$. Finally, since $\partial l_i/\partial \lambda_i = 1/\lambda_i$, the transformation $L = \log \Lambda$ has Jacobian $\mathcal{J}(L \to \Lambda) = |\Lambda|^{-1}$. Putting the three Jacobians together,
$$
\begin{aligned}
\mathcal{J}(Y \to X) &= \mathcal{J}\big(Y \to (V,L)\big) \cdot
\mathcal{J}\big((V,L) \to (V,\Lambda)\big) \cdot
\mathcal{J}\big((V,\Lambda) \to X \big) \\
&= \bigg( g(V) \prod_{i<j} (l_i - l_j) \bigg) |\Lambda|^{-1}
\bigg( g(V) \prod_{i<j} (\lambda_i - \lambda_j) \bigg)^{-1}
= \frac{1}{\lambda_1 \ldots \lambda_p}
\prod_{i<j} \frac{\log \lambda_i - \log \lambda_j}
{\lambda_i - \lambda_j}.
\end{aligned}
$$
The case $\lambda_i = \lambda_j$ is obtained by taking the limit of the fraction inside the product sign above as $\lambda_i \to \lambda_j$.

%-------------------------------------------------------------------
\subsection{Proof of Theorem \ref{thm:canon-avg-CLT}}

\subsubsection{Asymptotic normality}
By Remark 2.2 of \citet{Bhattacharya:2005}, Theorem 2.2 of \citet{Bhattacharya:2005} regarding the CLT for sample intrinsic means applies to $\Sym^+(p)$ with the canonical inner product because it has no cut-locus and the supremum of its sectional curvatures is nonpositive. Note that the measure $Q$ need not have compact support, as long as the second moment conditions in the statement of Theorem \ref{thm:canon-avg-CLT} are satisfied. Applying Theorem 2.2 of \citet{Bhattacharya:2005} implies that the conclusion of Theorem 2.1 of \citet{Bhattacharya:2005} regarding the CLT for sample Fr\'{e}chet means holds, and we state it next.

Rather than working with the variables $X_i$, it is convenient to work with the ``centered'' variables $X_i^I = M^{-1/2} X_i M^{-1/2}$. By equivariance of the canonical geometric mean and average with respect to the group action of $GL(p)$ (Section \ref{sec:canonical-mean}) with $G = M^{-1/2}$, the population canonical geometric mean of the $X_i^I$'s is the identity matrix $I$ (hence the superscript), and the canonical geometric average is $\ring{X}^I = M^{-1/2} \ring{X} M^{-1/2}$. To apply the CLT result above to the centered variables, let the distance $\rho$ needed there be the geodesic distance $d_{\C}$ \eqref{eq:canonical-distance} and let the chart $\phi(\cdot)$ be the Riemannian logarithmic map $\Log_{I}(\cdot) = \log(\cdot)$ \eqref{eq:Log} at $I$, which covers $\Sym^+(p)$ by \eqref{eq:diffeomorphism}. For two tangent vectors $U, \Theta \in \Sym(p)$ on the tangent space at $I$, the map $\phi$ induces the distance
\begin{equation}
\label{eq:rho-phi}
\rho^\phi(U,\Theta) = \rho\left(\phi^{-1}(U), \phi^{-1}(\Theta)\right)
= d_{\rm C}\left(e^U, e^\Theta\right),
\qquad U, \Theta \in \Sym(p)
\end{equation}
Then the image measure $Q^\phi$ has Fr\'{e}chet mean $M^\phi = \phi(M^I) = 0$ with respect to this distance. Similarly, $\hat{M}^\phi_{\rm C} = \phi(\ring{X}^I)$ is the Fr\'{e}chet mean of the image empirical measure $\hat{Q}^\phi_n = (1/n) \sum_{i=1}^n \delta(X^\phi_i)$, where
\[
X^\phi_i = \phi(X_i^I) = \log\left(M^{-1/2} X_i M^{-1/2}\right) = Y_i
\]
is assumed to have finite covariance. Further, from \eqref{eq:rho-phi}, define the half squared distance map
\begin{equation}
\label{eq:rho-phi-2}
\Theta \to G(U; \Theta) = \frac{1}{2}(\rho^\phi)^2(U,\Theta) = \frac{1}{2} d_{\rm C}^2\left(e^U, e^\Theta\right)
\end{equation}
together with its gradient and Hessian at $\Theta$
\begin{equation}
\label{eq:gradient-hessian}
\Psi(U; \Theta) = \nabla_\Theta G(U; \Theta) \in \Sym(p), \qquad
H(U; \Theta) = \partial_\Theta \vecd[\Psi(U; \Theta)] \in GL(q).
\end{equation}
Note the factor 1/2 in front of \eqref{eq:rho-phi-2}, missing in \citet[p. 1229]{Bhattacharya:2005}.
Both the gradient and Hessian exist because the map \eqref{eq:rho-phi-2} is at least twice differentiable. Then Theorem 2.1 of \citet{Bhattacharya:2005} states that
\begin{equation*}
\label{eq:CLT-intrinsic-1}
\sqrt{n}(\hat{M}^\phi_{\rm C} - M^\phi_{\rm C}) = \sqrt{n}\left[\log(M^{-1/2} \ring{X} M^{-1/2}) - 0\right] \Rightarrow N(0, K^{-1} C K^{-1}),
\end{equation*}
where
\begin{equation}
\label{eq:C-Lambda}
C = \Cov\left\{\vecd[\Psi(X_1^\phi; 0)]\right\} \in \Sym^+(q), \qquad
K = \E\left[H(X_1^\phi; 0)\right] \in GL(q),
\end{equation}
provided that $K$ is invertible. The required quantities $C$ and $K$ are computed in detail, next.

%-------------------------------------------------------------------
\subsubsection{Computation of the covariance $C$}
To compute the gradient in \eqref{eq:gradient-hessian}, we follow the method of \citet{Moakher:2005}. Let $\Theta + t A$ with $t \in \R$ and $A \in \Sym(p)$ be a straight line in the tangent space at $I$. From \eqref{eq:rho-phi-2}, define the scalar function
\begin{equation}
\label{eq:h}
h(t) = G(U,\Theta + t A) = \frac{1}{2} d_{\rm C}^2\left(e^U, e^{\Theta + t A}\right)
= \frac{1}{2} \tr\left[\log^2\left(e^{-U} e^{\Theta + t A}\right)\right]
\end{equation}
where the last expression was obtained using \eqref{eq:Exp} and \eqref{eq:canonical-distance}. Then the gradient in \eqref{eq:gradient-hessian} is the unique tangent vector $\Psi(U; \Theta)$ such that
\begin{equation}
\label{eq:gradient-def}
\left.\frac{dh(t)}{dt}\right|_{t=0} = \Langle \Psi(U; \Theta), A \Rangle_I
= \tr\left(\Psi(U; \Theta) A \right),
\end{equation}
where the Riemannian inner product \eqref{eq:canonical-inner-prod} on the tangent space at $I$ reduces to the Frobenius inner product \eqref{eq:Frobenius-inner-prod}.

Using the chain rule formula \citep[Prop. 2.1]{Moakher:2005}
\[
\frac{d}{dt} \tr\left[\log^2 X(t)\right] = 2 \tr\left[ \log X(t) X^{-1}(t) \frac{d}{dt} X(t)\right]
\]
in \eqref{eq:h} with $X(t) = e^{-U} e^{\Theta + t A}$ gives that the derivative of $h(t)$ at 0 is
\begin{equation}
\label{eq:h-deriv}
\left.\frac{dh(t)}{dt}\right|_{t=0} = \tr\left[\log\left(e^{-U} e^{\Theta}\right) \cdot e^{-\Theta} \left.\frac{d}{dt}e^{\Theta + t A}\right|_{t=0} \right].
\end{equation}
The first term inside the brackets has a polynomial expansion given by Goldberg's formula \eqref{eq:Goldberg}. Keeping terms only up to first order in $\Theta$,
\[
\begin{aligned}
\log\left(e^{-U} e^{\Theta}\right) &=
-U + \Theta
+ \frac{1}{2}\left(-U \Theta + \Theta U \right)
+ \frac{1}{12}\left(U^2 \Theta + \Theta U^2\right) - \frac{1}{6} U \Theta U \\
&- \frac{1}{720}\left(U^4 \Theta + \Theta U^4\right) + \frac{1}{180} \left(U \Theta U^3 + U^3 \Theta U\right) - \frac{1}{120}U^2 \Theta U^2 + \ldots
\end{aligned}
\]
Similarly, keeping terms only up to first order in $\Theta$ and $t$ in the second term in the brackets in \eqref{eq:h-deriv},
\[
%\begin{aligned}
e^{-\Theta} \left.\frac{d}{dt}e^{\Theta + t A}\right|_{t=0} =
(I - \Theta + \ldots) \frac{d}{dt}\left[I + \Theta + t A + \frac{1}{2}\left(\Theta + t A\right)^2 + \ldots\right]_{t=0}
= A + \frac{1}{2}\left(A \Theta - \Theta A\right) + \ldots
%\end{aligned}
\]
Replacing in \eqref{eq:h-deriv},
\[
\begin{aligned}
\left.\frac{dh(t)}{dt}\right|_{t=0} =& \tr\bigg\{\bigg[-U + \Theta + \frac{1}{12}\left(U^2 \Theta - 2 U \Theta U + \Theta U^2\right) \\
&+ \frac{1}{720}\left(U^4 \Theta - 4 U^3 \Theta U + 6 U^2 \Theta U^2 - 4 U \Theta U^3 + \Theta U^4\right) + \ldots
\bigg] A\bigg\}.
\end{aligned}
\]
Therefore, comparing with \eqref{eq:gradient-def}, we have that the desired gradient is
\begin{equation}
\label{eq:Psi}
\begin{aligned}
\Psi(U; \Theta) =& -U + \Theta + \frac{1}{12}\left(U^2 \Theta - 2 U \Theta U + \Theta U^2\right) \\
&+ \frac{1}{720}\left(U^4 \Theta - 4 U^3 \Theta U + 6 U^2 \Theta U^2 - 4 U \Theta U^3 + \Theta U^4\right) + \ldots
\end{aligned}
\end{equation}
which is indeed in $\Sym(p)$. In particular, $\Psi(X_1^\phi; 0) = - X_1^\phi = - Y_1$, with covariance $C = \Cov(\vecd(Y_1)) = \Sigma$.

%-------------------------------------------------------------------
\subsubsection{Computation of the expected Hessian $K$}
To obtain the Hessian in \eqref{eq:gradient-hessian}, we first vectorize the gradient \eqref{eq:Psi} as
\[
\begin{aligned}
\vecd[\Psi(U; \Theta)] =& -\vecd(U) + \vecd(\Theta) \\
&+ D'\left[\frac{1}{12}(U \otimes I - I \otimes U)^2 + \frac{1}{720}(U \otimes I - I \otimes U)^4 + \ldots\right] D \vecd(\Theta)
\end{aligned}
\]
where we have used property \eqref{eq:D} and the property that $\vec(AXB) = (B' \otimes A) \vec(X)$. Thus
\[
H(U; \Theta) = \frac{\partial \vecd[\Psi(U; \Theta)]}{\partial[\vecd(\Theta)]'} = I + D'\left[\frac{1}{12}(U \ominus U)^2 + \frac{1}{720}(U \ominus U)^4 + \ldots\right] D
\]
Finally, by \eqref{eq:C-Lambda} with $X_1^\phi = Y_1$, the desired matrix $K$ is as given in \eqref{eq:K}. Note that $K$ is PD and thus invertible.

%===================================================================
\bibliographystyle{apa}
%\bibliography{\bibpath/AS,\bibpath/MVSTATS,\bibpath/Neuroimaging,\bibpath/FDR}

\end{document}